\documentclass{article}

\usepackage[utf8]{inputenc}
\usepackage[margin=1in]{geometry}
\usepackage[titletoc,title]{appendix}

\usepackage{amsmath,amsfonts,amssymb,mathtools}

\usepackage{graphicx,float}

\usepackage{algpseudocode}
\usepackage{algorithm}

\usepackage{hyperref}
\usepackage{biblatex}
\title{Solving Inverse Problems with Hybrid Deep Image Priors: the challenge of preventing overfitting}
\author{Zhaodong Sun\thanks{This work was done as a master project at
École Polytechnique Fédérale de Lausanne in 2020 spring. } \\ \href{mailto:zhaodongsun@gmail.com}{zhaodongsun@gmail.com} \\ \\
Thesis Supervisors: Thomas Sanchez, Fabian Latorre, and Volkan Cevher}
\date{}

\begin{document}

\maketitle

\begin{abstract}
We mainly analyze and solve the overfitting problem of deep image prior (DIP). Deep image prior can solve inverse problems such as super-resolution, inpainting and denoising. The main advantage of DIP over other deep learning approaches is that it does not need access to a large dataset. However, due to the large number of parameters of the neural network and noisy data, DIP overfits to the noise in the image as the number of iterations grows. In the thesis, we use hybrid deep image priors to avoid overfitting. The hybrid priors are to combine DIP with an explicit prior such as total variation or with an implicit prior such as a denoising algorithm. We use the alternating direction method-of-multipliers (ADMM) to incorporate the new prior and try different forms of ADMM to avoid extra computation caused by the inner loop of ADMM steps. We also study the relation between the dynamics of gradient descent, and the overfitting phenomenon. The numerical results show the hybrid priors play an important role in preventing overfitting.  Besides, we try to fit the image along some directions and find this method can reduce overfitting when the noise level is large. When the noise level is small, it does not considerably reduce the overfitting problem.
\end{abstract}

\section{Introduction}

Classical compressed sensing (CS) methods are effective for sparse signal recovery and image reconstruction. CS utilizes the sparsity of some signals and can recover signals from much fewer measurements than the Nyquist sampling requirement. CS uses a sparse priors such as L1 norm to enforce the sparsity of a signal in other domains such as Fourier domain or wavelet domain. For anisotropic total variation (TV) norm, it is the L1 norm of gradient, which assumes the gradient of a signal is sparse. However, for images, these priors are not enough to capture the prior distribution, which causes low reconstruction performance from very few measurements.
\\
\\
As the development of generative adversarial network (GAN) \cite{goodfellow2014generative}, a deep learning method called compressed sensing using generative models (CSGM) \cite{bora2017compressed} uses a trained GAN as a prior. The priors in classical CS restrict the signal or image to the sparse domain, and cannot capture the real data distribution. However, a well trained GAN can get the distribution of a given dataset, which provides a more accurate prior for a specific task. Compared with classical methods, CSGM can achieve better reconstruction with the same measurements. When the number of measurements is very small , CSGM can still work while the classical methods fail to reconstruct an image.
\\
\\
However, CSGM also has some drawbacks. It needs a large dataset to train a GAN to capture the dataset distribution as the prior distibution. Classical CS does not need to access a dataset for model training, which is suitable when there is no large dataset available. Besides,  it is not easy to train a GAN due to minmax loss function and mode collapse. It might not converge with minmax optimization, and the trained GAN with mode collapse might produce very similar images, which is not a good approximation of the data distribution. Even if we have a trained GAN, it takes hundreds of iterations and several restart to get a good reconstruction. \cite{bojanowski2017optimizing}
\\
\\
Deep image prior (DIP) \cite{ulyanov2018deep} provides a new option similar to CSGM. It uses a random initialized convolutional neural network as the prior, which does not need a dataset for neural network training. The reconstruction results are comparable with some state-of-the-art data-driven methods \cite{lai2017deep, ledig2017photo} which need a large dataset to train a neural network. However, we still need to optimize over a large number of weights in a neural network for each single image, and it takes several minutes on GPU for one image. Besides, the network is over-parameterized, so it will easily overfit to the noisy measurements. The authors of DIP propose to use early stopping to avoid overfitting, however, it is not clear when to stop learning.
\\
\\
We will describe the overfitting phenomenon of DIP in this paragraph. The overfitting phenomenon of DIP can be observed from the learning curves and reconstructed images. From learning curves, the peak signal noise ratio (PSNR) will first reach a peak and then gradually decrease during training. The curve is shown at Figure \ref{fig:all_alg_compare_sub1} with DIP legend. This curve indicates that the reconstructed image will first approach the true image. As the number of iterations grows, the reconstructed image gradually deviate from the true image and approach the noisy image. From a reconstructed image in the first column of Figure \ref{fig:celeba_img_result}, we can observe, the image have a good reconstruction at a small iteration and will gradually be noisy. The noisy image means it deviates from the true image and approach the noisy measurements.
\\
\\
In this thesis, we try to incorporate deep image prior with another prior to avoid overfitting. Since the new prior is performed on the output of the neural network, the variable in the original problem can be decomposed into two variables. One is the reconstructed image and another one is the weights of the neural network. The constraint is that the reconstructed image equals the output of the neural network. The exact formulation of the problem is at equation \ref{eq:ori_obj_with_prior_1}. For this kind of optimization problem, we could use alternating direction method-of-multiplier (ADMM) \cite{boyd2011distributed}. We propose two types of ADMM algorithms to test the effectiveness of the prior. The advantage of our proposed ADMM algorithms is that it can use a denoising algorithm as a prior.This can be done by replacing the proximal operator in one of ADMM steps with a denoising algorithm. One of our proposed ADMM algorithms can avoid inner loops to reduce computation. The numerical experiments show the combination of DIP and denoising algorithms can avoid overfitting to some extent. 
\\
\\
To know why DIP overfits, we analyzed the dynamics of gradient descent, when solving the optimization problem. What happens during training is that the model will fit the most important components of the image, and then it will start fitting the noise. To avoid this issue, we could only fit the image along some directions that mostly span the image. In chapter 5.5, we show this method is effective when the noise level is large.
\section{Background}
\subsection{Classical Compressed Sensing Methods}
For linear inverse problems, we obtain the partial measurement $b \in \mathbb{R}^m$ from a reference image $x \in \mathbb{R}^n$
\begin{equation}
b=Ax + \epsilon
\end{equation}
where $\epsilon \in \mathbb{R}^m$ is Gaussian noise with distribution $\mathcal{N}(0, \zeta^2I)$. $A$ is the measurement matrix with shape $m\times n$ ($m\ll n$). If $x$ is an image, the measurement matrix $A$ can be down sampling,  random pixel removal, or some other operations in specific cases. This is a dimension reduction operation and we need to solve an ill-posed problem. To solve the inverse problem, we consider to formulate the problem from Maximum a posteriori (MAP). The logarithm of the posterior distribution is 
\begin{equation}
\log p (x|b) \propto \log p(b|x) + \log p(x)
\end{equation}
For the likelihood $p(b|x)$, it is Gaussian distribution $\mathcal{N}(Ax, \zeta^2I)$. To maximize the posterior $p(x|b)$, the problem can be
\begin{equation}
\min_{x} \frac{1}{2}\lVert b-Ax \rVert_2^2 + R(x)
\end{equation}The regularization term $R(x)$ is proportional to the log prior $\log p(x)$. Since images have sparse representation in some domains, $R(x)$ can be L1 norm $\lVert \Phi x \rVert_1$ where $\Phi$ could be discrete cosine transform or wavelet transform. In this case, this optimization problem with L1 sparse prior is called Lasso \cite{tibshirani1996regression}. Most images also have low variations in local pixels, so R(x) can also be TV norm \cite{Barbero11}. Anisotropic TV norm \cite{JMLR:v19:13-538} is the L1 norm of the signal gradient and $\Phi$ can be regarded as a gradient operator\footnote{In the following chapters, the TV norm will be anisotropic, which is defined as  $\lVert x \rVert _{TV}=\sum_{i,j}\lvert x_{i,j}-x_{i+1,j} \rvert + \lvert x_{i,j}-x_{i,j+1} \rvert$}. This prior assumes the sparsity of the gradient, and minimizing the TV norm can make sure that local pixels do not change too much
\\
\\
The above problem often has a convex but non-smooth $R(x)$, so we can use proximal gradient methods such as iterative soft thresholding algorithm (ISTA) \cite{daubechies2004iterative} and fast iterative soft thresholding algorithm (FISTA) \cite{beck2009fast} to solve the optimization problem above. If $k$ is the iteration, ISTA has the convergence rate $\mathcal{O}(1/k)$ while FISTA has a faster convergence rate $\mathcal{O}(1/k^2)$. For the basic proximal gradient methods like ISTA\cite{daubechies2004iterative}, the update step is 
\begin{equation}
x_{t+1}=\text{prox}_{\beta R}(x_t-\beta \nabla_x f(x)|_{x=x_t})
\end{equation}where $\text{prox}_{g}(y)=\operatorname*{argmin}_x \frac{1}{2}\lVert y-x \rVert_2^2 + g(x)$ , $f(x)=\frac{1}{2}\lVert b - Ax \rVert_2^2$, and $\beta=\frac{1}{\lVert A \rVert^2}$
\subsection{Deep learning methods for compressed sensing}
A direct method is to train a neural network $h_{\phi}(b):\mathbb{R}^m \to \mathbb{R}^n$ with the measurement $b$ as the input and the reconstructed image as the output \cite{adler2018deep}. This is a data-driven method, which requires a dataset. Suppose we have a dataset with pairs of measurements $b$ and true images $x$, and they have the distribution $(b, x) \sim \mathcal{D}$. The optimization problem to train the neural network is
\begin{equation}
\min_{\phi} \mathbb{E}_{(b, x) \sim \mathcal{D}} \big[ \lVert h_{\phi}(b) - x \rVert_2^2 \big]
\end{equation}
When we want to solve the inverse problem with a new measurement, we just use this measurement as the input of the trained neural network $h_{\phi}(b)$ and the output is the reconstructed image. However, this model does not utilize the measurement matrix $A$ and largely depends on the dataset to infer the true signal from the measurement. When the measurement matrix changes, we may train the neural network again.
\\
\\
The classical compressed sensing problem can also be combined with deep neural networks. We can first use a dataset to train a generative adversarial network (GAN) \cite{goodfellow2014generative} $G(z;\theta): \mathbb{R}^p \to \mathbb{R}^n$ with $z \in \mathbb{R}^p$ as the latent variable and $\theta$ is the network weights. To train a GAN, we have to solve a minmax optimization problem.
\begin{equation}
\min_{\theta} \max_{\phi} \mathbb{E}_{x \sim p_{data}(x)}[\log{D(x; \phi)}]+\mathbb{E}_{z \sim p_z(z)}[\log{\big( 1-D(G(z;\theta);\phi) \big)}]
\end{equation}where $G(z;\theta)$ is the generator and $D(x;\phi): \mathbb{R}^n \to [0,1]$ is a discriminator. When we maximize the loss function over the weights $\phi$ of the discriminator, we try to achieve the following things. When the input of the discriminator is a true sample from the distribution $p_{data}(x)$, the output will be large. When the input of the discriminator is a fake sample from $G(z;\theta)$, the discriminator will be low. When we minimize the loss function over the weights $\theta$, we try to fool the discriminator and encourage the generator to generate images similar to the real images.
\\
\\
To find the solution of the minmax optimization problem, this internal maximization problem can be solved and the above optimization problem can be transformed to
\begin{equation}
\min_{\theta} -\log(4)+ 2 \cdot\text{JSD}(p_{data}||p_{G(z;\theta)})
\end{equation}where $\text{JSD}(p_1||p_2)= \frac{1}{2}\text{KL}(p_1||\frac{p_1+p_2}{2}) + \frac{1}{2} \text{KL}(p_2||\frac{p_1+p_2}{2})$ is Jensen–Shannon divergence and $\text{KL}$ is Kullback-Leibler divergence. Both JSD and KL are distance metrics between two probability distributions. $p_{G(z;\theta)}$ is the distribution of the generator outputs. The distribution of $z$ can an arbitrary distribution, but in practice, it is usually Gaussian distribution or uniform distribution. Since JSD is non-negative, and zero when $p_1=p_2$, the minimal loss value is $-\log(4)$ when $p_{data}=p_{G(z;\theta^*)}$. This means the trained generator can model the distribution of the dataset and can be used as a prior in the compressed sensing problems.
\\
\\
Since the generator $G(z;\theta^*)$ captures the prior distribution $p_{data}(x)$, we can use this generator as a kind of prior \cite{bora2017compressed}\cite{patel2019bayesian}. The generator can only generate images from the prior distribution, so we can set $x=G(z;\theta^*)$ as a constraint which is another form of prior. The optimization problem will be
\begin{equation}
\begin{split}
\min_{x, z} &\frac{1}{2}\lVert b-Ax \rVert_2^2 \\
&\text{s.t.} \  x = G(z;\theta^*)
\end{split}
\end{equation}
This method uses a trained generator as a prior, which can achieve better reconstruction performance with fewer measurements, compared with TV and L1 priors. However, this method needs some data to train the generator and it is not easy to train a GAN \cite{goodfellow2014generative}. Due to the minmax optimization, GAN might fail to convergence. A GAN with Mode collapse can only generate very similar images or even same images, which cannot represent the distribution of the dataset. Note that when we train the GAN $G(z; \theta)$, the latent variable $z$ also has a prior distribution such as Gaussian distribution, so CSGM \cite{bora2017compressed} also incorporates the Gaussian prior of $z$. The new optimization problem will be
\begin{equation}
\begin{split}
\min_{x, z} &\frac{1}{2}\lVert b-Ax \rVert_2^2 + \lambda \lVert z \rVert_2^2\\
&\text{s.t.} \  x = G(z;\theta^*)
\end{split}
\end{equation}
\subsection{Deep image prior for compressed sensing}
\subsubsection{Introduction to deep image prior}
 Deep image prior (DIP) \cite{ulyanov2018deep} uses a random initialized generator. The problem is
\begin{equation}
\begin{split}
\min_{x, \theta} &\frac{1}{2}\lVert b-Ax \rVert_2^2 \\
&\text{s.t.} \ x = G(z;\theta)
\end{split}
\end{equation}
This new problem is very similar to CSGM with a trained generator. However, CSGM requires the trained generator $G(z;\theta^*)$ and optimizes over the latent variable $z$. The deep image prior problem only needs a random initialized generator. It optimizes over the weights of the generator while the latent variable $z$ is kept constant. Deep image prior utilizes the convolutional structure of the neural network as the prior. The reconstructed image should fall in the space of the generator output. The generator in DIP has more flexibility than CSGM since the weights of generator varies during optimization.The advantage of deep image prior is that it does not need a dataset, but it will easily overfit due to the large number of trainable weights. However, for the trained generator $G(z;\theta^*)$ in CSGM, we only need to optimize over the latent variable $z$ and there is no overfitting phenomenon in the numerical experiments.
\subsubsection{Improvement of deep image prior}
Since deep image prior will overfit easily, there are some works about the generalization of over-parameterized networks. In \cite{heckel2019denoising}, they analyze why deep image prior can fit image faster than noise and demonstrate early stopping is necessary. They mainly study the dynamics of gradient descent and find the Jacobian of the network output with respect to the weights plays an important role. They get the singular values and vectors of the Jacobian. The image can be spanned by a subset of singular vectors corresponding to the largest values while the noise can be spanned by all the singular vectors. Since the singular values determine the fitting speed at the corresponding singular vectors, the image is fitted faster than noise. They provide singular vectors and values for 1D signals and test the singular values and vectors can be regarded as constant for over-parameterized networks during optimization. In chapter 5 of this thesis, we will give more results related to the Jacobian for images and use a new method to avoid overfitting. \cite{oymak2019generalization} also analyzes the overfitting of an over-parameterized neural network. They divide the space spanned by the singular vectors of the Jacobian into information space and nuisance space. The information space corresponds to the space spanned by the first several singular vectors and the nuisance space is the left space.
\\
\\
Some papers also propose new priors combined with deep image prior to achieve better performance. \cite{liu2019image} added a TV prior in the loss function and the numerical experiment shows it can improve the reconstruction results. The optimization problem for \cite{liu2019image} is
\begin{equation}
\begin{split}
\min_{x, \theta} &\frac{1}{2}\lVert b-Ax \rVert_2^2 +\lambda \lVert x \rVert_{TV} \\
&\text{s.t.} \  x = G(z;\theta)
\end{split}
\end{equation}
However, they directly use gradient descent to optimize the loss function. The TV norm is non-smooth and some proximal gradient methods should be used to optimize the loss function. In this thesis, we propose ADMM algorithms to tackle this problem. Moreover, our proposed methods can also be used with denoising algorithms as priors.
\\
\\
DeepRED \cite{mataev2019deepred} combines denoising algorithms with DIP. This work uses the technique in regularization as denoising (RED) \cite{romano2017little} to add a denoising algorithm. It can also calculate the gradient through the denoising algorithm. The optimization problem for deepRED \cite{mataev2019deepred} is similar to the previous one.
\begin{equation}
\begin{split}
\min_{x, \theta} &\frac{1}{2}\lVert b-Ax \rVert_2^2 +\frac{\lambda}{2} x^T(x-f(x)) \\
&s.t. \  x = G(z;\theta)
\end{split}
\end{equation}
Where $f(x)$ is the denoising algorithm. They use ADMM algorithm to solve the optimization problem above. In their ADMM algorithm, they have an inner loop in one of the ADMM steps while one of our proposed ADMM algorithms does not have any inner loop. Since the inner loops will have to evaluate the denoising algorithm several times in one iteration, this will add computation burden.
\\
\\
Recently, DIP with learned regularization \cite{van2018compressed} is also proposed. It has a prior on the network weights. Their optimization problem is
\begin{equation}
\begin{split}
\min_{x, \theta} &\frac{1}{2}\lVert b-Ax \rVert_2^2 + \lambda_{TV}\lVert x \rVert_{TV} + \lambda_L (\theta - \mu)^T \Sigma^{-1} (\theta - \mu) \\
&s.t. \  x = G(z;\theta)
\end{split}
\end{equation}They add a TV prior like \cite{liu2019image} and a quadratic term induced from Gaussian distribution $\theta \sim \mathcal{N}(\mu, \Sigma)$. When $\mu$ is zero and $\Sigma$ is identity matrix, the new prior is degraded to L-2 regularization. However, they need a small dataset to get the learned $\mu$ and $\Sigma$ and the high dimensions will make it infeasible to learn $\Sigma$. Therefore, they simplify the covariance matrix. They assume that the parameters within the layer $l \in \{1,2, ..., L\}$  are independent, so they can simplify the covariance matrix to a $L \times L$ matrix and the mean vector to a vector with length L. When they want the full mean and covariance matrix, they can just expand the simplified matrix and vector with replicated values.
\\
\\
The mean and covariance can be obtained from samples of learned weights. They run the the basic DIP optimization $\min_{\theta} \frac{1}{2}\lVert b_i-AG(z;\theta) \rVert_2^2$ for each measurement $b_i$ in the dataset $\mathcal{S} = \{b_1, b_2, ..., b_Q\}$ and get the corresponding weights $\mathcal{W}=\{w_1, w_2, ..., w_Q\}$. They do statistics on the weights $\mathcal{W}$ to get the simplified mean and covariance matrix. 
\\
\\
They show the result is improved compared with only DIP. This method still requires the dataset for the learned regularization, but it only requires the measurements rather than the ground truth. However, they do not compare the performances of the L-2 regularization and the learned regularization. Also, they use a TV norm as another prior, and it is not clear how much the performance is influenced by the TV norm.
\subsubsection{Methods for preventing overfitting of deep image prior}
One method to prevent overfitting is to use probabilistic distribution sampling to get the posterior distribution, which is not point estimation for maximum a posteriori. \cite{cheng2019bayesian} uses stochastic gradient Langevin dynamics (SGLD) \cite{welling2011bayesian} update. SGLD is a Markov chain Monte Carlo (MCMC) sampler from stochastic gradient descent by adding Gaussian noise to the gradient and can provide posterior samples for Bayesian inference. Bayesian posterior sampling is effective to prevent overfitting. The loss function from logarithm posterior distribution and the gradient update are
\begin{equation}
\begin{split}
L(\theta) = \frac{1}{2}\lVert b-AG(z;\theta) \rVert_2^2 + \lambda \lVert \theta \rVert_2^2
\\
\theta_{t+1} = \theta_t - \frac{\epsilon}{2}\nabla_{\theta}L(\theta)|_{\theta=\theta_t}+\eta_t
\\
\eta_t \sim \mathcal{N}(0, \epsilon)
\end{split}
\end{equation}where $\epsilon$ is the stepsize. The first and second term in the loss function are proportional to the log-likelihood and the logarithm prior distribution of the network weights, respectively. This update is similar to gradient descent except that it adds noise to each update result. Their results show this update can avoid overfitting and also has better reconstruction than gradient descent.
\\
\\
Another method is to modify the network architecture $G(z;\theta)$. One such approach is the Deep decoder architecture \cite{heckel2018deep}, which only has 1x1 convolutions and upsampling operators. For DIP, it uses a U-net-like \cite{ronneberger2015u} architecture including 3 by 3 convolution, downsampling and upsampling. Only $1 \times 1$ convolution is a non-convolution network, and it doesn't combine the local information. However, deep decoder also has upsampling operation such as bilinear upsampling, which can combine the local information. They also prove that deep decoder can only fit a small portion of noise from theory and numerical results.
\\
\\
Another paper \cite{metzler2018unsupervised} uses Stein unbiased risk estimator (SURE) \cite{stein1981estimation} as the loss function. For denoising problem, the mesurement $b$ is obtained from $b = x + w$ where $w \sim \mathcal{N}(0, \sigma^2I)$. SURE is the expectation of mean squared error between the reconstructed image $\hat{x}$ and the true image $x$. 
\begin{equation}
\text{SURE}(\hat{x}) = \mathbb{E}_{w \sim \mathcal{N}(0, \sigma^2I)}\Big[\frac{1}{N}\lVert x - \hat{x} \rVert_2^2\Big]
\end{equation}
where N is the number of pixels. If we replace the reconstructed image $\hat{x}$ with the network output $G(b; \theta)$, and use Stein's lemma, it becomes
\begin{equation}
\mathbb{E}_{w \sim \mathcal{N}(0, \sigma^2I)}\Big[\frac{1}{N}\lVert x - G(b; \theta) \rVert_2^2\Big]
=\frac{1}{N}\lVert b - G(b; \theta) \rVert_2^2-\sigma^2+ \frac{2\sigma^2}{N}\mathbf{Tr}(\frac{\partial G(b; \theta)}{\partial b})
\end{equation}
Therefore, we can use this loss function as the objective, which is the generalization error.
\begin{equation}
\min_{\theta} \frac{1}{2} \lVert b - G(b; \theta) \rVert_2^2+ \sigma^2 \mathbf{Tr}(\frac{\partial G(b; \theta)}{\partial b})
\end{equation}
SURE adds a new term, which is the trace of the Jacobian of network output with respect to the input. Since the input and output of the network is very large, it is expensive to get the trace of the Jacobian. Also, we have to calculate the gradient of the trace of Jacobian, which is computationally infeasible for large images. Therefore, they use Monte-Carlo SURE \cite{ramani2008monte} to approximate the trace of the Jacobian. This method can be effective to prevent overfitting, but it could cause instability of training maybe due to the Monte-Carlo approximation of SURE. When the training iteration is small ,it shows a good effect to prevent overfitting, however, the loss will fluctuate for large iterations.
\subsection{Using denoising algorithms as priors}
Denoising algorithms can also be regarded as regularizers. For example, the proximal operator of total variation $\text{prox}_{TV}(y)=\operatorname*{argmin}_x \frac{1}{2}\lVert y-x \rVert_2^2 + \lVert x \rVert_{TV}$ can work as a denoising algorithm, which means it is possible for a denoising algorithm as a prior even though most denoising algorithms do not have an explicit prior form like TV norm.
\subsubsection{Plug-and-play prior}
Plug-and-play prior \cite{venkatakrishnan2013plug} is a method to incorporate denoising algorithms as priors into a compressed sensing problem. The optimization problem is shown below.
\begin{equation}
\begin{split}
\operatorname*{min}_{y, x} 
\frac{1}{2}
\lVert b-Ay \rVert ^2_2
+
R(x)
\\
\text{s.t.} \ x=y
\end{split}
\end{equation}
Plug-and-play prior splits the optimization variable into two variables $x$ and $y$, and adds a constraint to make them equal. They propose to use ADMM algorithm to solve the problem. The augmented Lagrangian is
\begin{equation}
\mathcal{L}_{\rho}(x, y, u)=
\frac{1}{2}
\lVert b-Ay \rVert ^2_2
+ R(x) +
\frac{\rho}{2}
\lVert x-y+u \rVert_2^2
- \frac{\rho}{2} \lVert u \rVert_2^2
\label{eq:lag_3}
\end{equation}

The ADMM steps can be obtained from the minmax on the augmented Lagrangian $\text{min}_{x, y}\text{max}_u \mathcal{L}_{\rho}(x, y, u)$
\begin{equation}
\begin{split}
x_{t+1} &= \operatorname*{argmin}_{x} R(x)+\frac{\rho}{2}\lVert x-(y_t-u_t) \rVert_2^2
=\text{prox}_{\frac{R}{\rho}}(y_t-u_t)
\\
y_{t+1} &= \operatorname*{argmin}_{y} \frac{1}{2} \lVert b-Ay\rVert ^2_2 + \frac{\rho}{2} \lVert x_{t+1}-y+u_t\rVert_2^2
\\
u_{t+1} &= \operatorname*{argmax}_{u} \frac{\rho}{2}
\lVert x_{t+1}-y_{t+1}+u \rVert_2^2
- \frac{\rho}{2} \lVert u \rVert_2^2
\end{split}
\label{eq:admm_detail_3}
\end{equation}

The minimization on x in the ADMM steps above can be formulated as a proximal operator. This proximal operator can be replaced with a denoising algorithm and the numerical experiments show it can work with some state-of-the-art denoising algorithms.
\subsubsection{Regularization by Denoising (RED)}
RED \cite{mataev2019deepred} gives an explicit prior with a denoising algorithm. The optimization problem is
\begin{equation}
\operatorname*{min}_{y, x} 
\frac{1}{2}
\lVert b-Ax \rVert ^2_2
+
\frac{\lambda }{2} x^T\big[ x - f(x) \big]
\end{equation}The prior is $\rho(x) = \frac{1 }{2} x^T\big[ x - f(x) \big]$ where $f(x)$ is a denoising algorithm. It calculates the cross correlation between $x$ and the residual $x-f(x)$. When $x$ is an image without noise, which means $x \approx f(x)$, the prior value will be almost zero. However, when $x$ is a noisy image, the prior $\rho(x)$ value will be non-zero.
\\
\\
The problem of this prior is that we cannot get the gradient of $f(x)$ since it is a black box function. The authors of RED \cite{romano2017little} assume that a denoising algorithm should have \textit{local homogeneity}. The property is that $f(cx)=cf(x)$ and $|c-1|<\epsilon$ for a very small $\epsilon$. This property also induces $f(x)=\nabla_x f(x)x$. This property holds for almost all denoising algorithms \cite{mataev2019deepred}, so we can get the gradient of the prior
\begin{equation}
\nabla_x \rho(x)=\frac{1}{2}\big[ 2x-f(x)-\nabla f(x)x\big]=x-f(x)
\end{equation}
which means we don't need to know the gradient of the denoising algorithm.
\section{Methods}

\subsection{Problem formulation}
The form of the deep image prior (DIP) is an untrained convolutional neural network (CNN). The objective can be formulated as \ref{eq:ori_obj} where $b \in \mathbb{R}^m$ is the noisy measurement from $b=Ax^\natural+\epsilon$, $\epsilon$ is Gaussian noise with the distribution $\mathcal{N}(\mathbf{0}, \sigma^2 \mathbf{I})$, and $A\in\mathbb{R}^{m\times n}$ is a measurement matrix. For convenience, we directly use the flattened image vector $x\in\mathbb{R}^n$. Besides, $x^{\natural} \in \mathbb{R}^n$ is the true image. DIP is not like L1 or L2 regularization, it adds a constraint $x=G(\theta)$ to let the solution fall in the space of the CNN output, 
\begin{equation}
\begin{split}
\operatorname*{min}_{\theta, x} 
\frac{1}{2}
\lVert b-Ax \rVert ^2_2
\\
\text{s.t.} \ x=G(\theta) 
\end{split}
\label{eq:ori_obj}
\end{equation}
Actually, the CNN should be $G(z;\theta)$ where the weights are $\theta\in\mathbb{R}^w$ and the input is $z\in\mathbb{R}^n$. Since $z$ is randomly initialized with standard Gaussian distribution and is kept constant during optimization\footnote{Note that $z$ can be perturbed by noise during optimization to achieve better results as described in the original DIP paper\cite{ulyanov2018deep}, but we use $z$ as a constant for convenience. Otherwise, the problem will be a stochastic optimization. The influence of the perturbation of z can be found at \cite{metzler2018unsupervised}, which is related to the Jacobian of the CNN}, we ignore $z$ and write the CNN as $G(\theta)$. 
\\
\\
we propose to add another prior, in the form of an additional regularizer term R(x) in the objective function. The prior can be explicit such as TV norm or L1 norm. The prior can also be implicit, which means we don't have the $R(x)$, but we have the proximal operator of $R(x)$, which is $\text{prox}_{R}(x)=\operatorname*{argmin}_{y} \frac{1}{2}\lVert y-x \rVert_2^2+R(y)$. The implicit case is applied to denoising algorithms as proximal operator. In the following sections, we can see the proximal operator is enough for ADMM steps. The proximal operator can also be interpreted as a denoising algorithm.
\\
\\
We will propose two options to incorporate R(x). We call them DIP-ADMM-v1 and DIP-ADMM-v2 in \ref{eq:ori_obj_with_prior_1} and \ref{eq:ori_obj_with_prior_2}, respectively. The formulation in \ref{eq:ori_obj_with_prior_1} is similar to \cite{gomez2019fast} except that we use an untrained neural network, and optimize over the weights while \cite{gomez2019fast} uses a trained neural network, and optimizes over the input of the network. In the next section, we will derive the ADMM steps and compare these two formulations.

\begin{equation}
\begin{split}
\operatorname*{min}_{\theta, x} 
\frac{1}{2}
\lVert b-Ax \rVert ^2_2
+
R(x)
\\
\text{s.t.} \ x=G(\theta) 
\end{split}
\label{eq:ori_obj_with_prior_1}
\end{equation}
\begin{equation}
\begin{split}
\operatorname*{min}_{\theta, x} 
\frac{1}{2}
\lVert b-AG(\theta) \rVert ^2_2
+
R(x)
\\
\text{s.t.} \ x=G(\theta) 
\end{split}
\label{eq:ori_obj_with_prior_2}
\end{equation}
\subsection{Deep image prior with a denoising algorithm by ADMM}
Problem \ref{eq:ori_obj_with_prior_1} and \ref{eq:ori_obj_with_prior_2} are equivalent to the problem represented by the augmented Lagarangian. 
\begin{equation}
\label{eq:lag_obj}
\operatorname*{min}_{\theta, x} \operatorname*{max}_{u} \mathcal{L_{\rho}}(x, \theta, u)
\end{equation}
\subsubsection{DIP-ADMM-v1}
In this subsection, we will introduce the ADMM algorithms for \ref{eq:ori_obj_with_prior_1}. First, the augmented Lagrangian with scaled dual variable $u$ is at \ref{eq:lag_1} 

\begin{equation}
\mathcal{L}_{\rho}^{(1)}(x, \theta, u)=
\frac{1}{2}
\lVert b-Ax \rVert ^2_2
+ R(x) +
\frac{\rho}{2}
\lVert x-G(\theta)+u \rVert_2^2
- \frac{\rho}{2} \lVert u \rVert_2^2
\label{eq:lag_1}
\end{equation}
The corresponding ADMM steps are at \ref{eq:admm_detail_1}. 
\begin{equation}
\begin{split}
x_{t+1} &= \operatorname*{argmin}_{x} \frac{1}{2}\lVert b-Ax \rVert ^2_2
+ R(x) +
\frac{\rho}{2}
\lVert x-G(\theta_t)+u_t \rVert_2^2
\\
\theta_{t+1} &= \operatorname*{argmin}_{\theta} \lVert x_{t+1}-G(\theta)+u_t \rVert_2^2
\\
u_{t+1} &= \operatorname*{argmax}_{u} \frac{\rho}{2}
\lVert x_{t+1}-G(\theta_{t+1})+u \rVert_2^2
- \frac{\rho}{2} \lVert u \rVert_2^2
\end{split}
\label{eq:admm_detail_1}
\end{equation}
In the first step in \ref{eq:admm_detail_1}, corresponding to exact minimization over the variable x, the objective is convex, so we can do ISTA or FISTA for some iterations to get an approximate solution. The iteration of ISTA for the update of variable $x$ is 
\begin{equation}
x_{t+1} = \text{prox}_{\frac{R}{\lVert A^TA\rVert+\rho}}(x_t-\frac{1}{\lVert A^TA\rVert+\rho}\nabla_x\mathcal{L}_{\rho}^{(1)}(x, \theta_t, u_t)|_{x=x_t})
\end{equation}
If $R(x)=0$ or $A=I$, there is also a closed form for exact minimization, but this is not applicable to our case. For the update of variable $\theta$, we can't get the exact minimization since it is nonconvex, so we could take one gradient descent step just as \cite{gomez2019fast} did in their primal updates.  For the update of variable $u$, we could take one gradient ascent step as \cite{boyd2011distributed}\cite{gomez2019fast} did in their dual ascent.
\subsubsection{DIP-ADMM-v2}
For \ref{eq:ori_obj_with_prior_2}, we still get the augmented Lagrangian at \ref{eq:lag_2}. The difference between this augmented Lagrangian and the previous one is in the first term. In DIP-ADMM-v1, the first term in the augmented Lagrangian is $\frac{1}{2} \lVert b-Ax \rVert ^2_2$, while the first term in DIP-ADMM-v2 is $\frac{1}{2}
\lVert b-AG(\theta) \rVert ^2_2$. Since we have the constraint $x = G(\theta)$, these two augmented Lagrangian functions are equivalent. However, this different formulation in DIP-ADMM-v2 causes different ADMM steps.
\begin{equation}
\mathcal{L}_{\rho}^{(2)}(x, \theta, u)=
\frac{1}{2}
\lVert b-AG(\theta) \rVert ^2_2
+ R(x) +
\frac{\rho}{2}
\lVert x-G(\theta)+u \rVert_2^2
- \frac{\rho}{2} \lVert u \rVert_2^2
\label{eq:lag_2}
\end{equation}
The corresponding ADMM steps are at \ref{eq:admm_detail_2}
\begin{equation}
\begin{split}
x_{t+1} &= \operatorname*{argmin}_{x} R(x)+\frac{\rho}{2}\lVert x-(G(\theta_t)-u_t) \rVert_2^2
=\text{prox}_{\frac{R}{\rho}}(G(\theta_t)-u_t)
\\
\theta_{t+1} &= \operatorname*{argmin}_{\theta} \frac{1}{2} \lVert b-AG(\theta) \rVert ^2_2 + \frac{\rho}{2} \lVert x_{t+1}-G(\theta)+u_t\rVert_2^2
\\
u_{t+1} &= \operatorname*{argmax}_{u} \frac{\rho}{2}
\lVert x_{t+1}-G(\theta_{t+1})+u \rVert_2^2
- \frac{\rho}{2} \lVert u \rVert_2^2
\end{split}
\label{eq:admm_detail_2}
\end{equation}
For the update of $x$, we can observe it is the form of proximal operator of $R(x)/\rho$ with input $G(\theta_t)-u_t$, so it is not necessary to take several ISTA iterations to get the exact minimization. For the update of $\theta$ and $u$, we still take one descent step and ascent step, respectively.
\\
\\
The main difference between DIP-ADMM-v1 and DIP-ADMM-v2 is the $x$ step. For DIP-ADMM-v1, the minimization in $x$ step needs some iterations and have to evaluate the proximal operator for several times. For DIP-ADMM-v2, it only has to do one step and evaluate the proximal operator only once. This could simplify the algorithm and decrease the evaluation times of the proximal operator since some proximal operators are expensive to evaluate.

\subsubsection{Plug-and-play prior}
Plug-and-play prior \cite{venkatakrishnan2013plug} is a classical method for model-based reconstruction. It is very similar to DIP-ADMM-v2 except that the neural network is replaced with a trainable variable $y$. The problem is
\begin{equation}
\begin{split}
\operatorname*{min}_{y, x} 
\frac{1}{2}
\lVert b-Ay \rVert ^2_2
+
R(x)
\\
\text{s.t.} \ x=y
\end{split}
\label{eq:obg_p3}
\end{equation}
Similarly, we write the augmented Lagrangian. 
\begin{equation}
\mathcal{L}_{\rho}^{(3)}(x, y, u)=
\frac{1}{2}
\lVert b-Ay \rVert ^2_2
+ R(x) +
\frac{\rho}{2}
\lVert x-y+u \rVert_2^2
- \frac{\rho}{2} \lVert u \rVert_2^2
\label{eq:lag_3}
\end{equation}
The ADMM steps are
\begin{equation}
\begin{split}
x_{t+1} &= \operatorname*{argmin}_{x} R(x)+\frac{\rho}{2}\lVert x-(y_t-u_t) \rVert_2^2
=\text{prox}_{\frac{R}{\rho}}(y_t-u_t)
\\
y_{t+1} &= \operatorname*{argmin}_{y} \frac{1}{2} \lVert b-Ay\rVert ^2_2 + \frac{\rho}{2} \lVert x_{t+1}-y+u_t\rVert_2^2
\\
u_{t+1} &= \operatorname*{argmax}_{u} \frac{\rho}{2}
\lVert x_{t+1}-y_{t+1}+u \rVert_2^2
- \frac{\rho}{2} \lVert u \rVert_2^2
\end{split}
\label{eq:admm_detail_3}
\end{equation}Both $x$ step and $y$ step can be minimized exactly and we just take a gradient ascent step for the update of variable $u$.
\subsubsection{Denoising algorithm as a proximal operator}

The two ADMM algorithms have to evaluate the proximal operator in $x$ step. In plug-and-play prior \cite{venkatakrishnan2013plug}, they replace the proximal operator with a denoising algorithm such as BM3D\cite{dabov2007image} so that the prior can be induced by any denoising algorithm. Here, we try to replace the proximal operator with a denoising algorithm to prevent the overfitting of deep image prior.
\\
\\
For TV prior, it has the explicit form, and the proximal operator of TV norm also works as a denoising algorithm. However, for other denoising algorithms like BM3D \cite{dabov2007image}, it doesn't have explicit prior form and only has the proximal operator. We can still use the prior information from denoising algorithms via the proximal operator.

\subsection{Implementation details}
The detailed procedures of DIP-ADMM-v1, DIP-ADMM-v2, and plug-and-play prior are shown in algorithms \ref{alg:DIP-ADMM-v1}, \ref{alg:DIP-ADMM-v2}, and \ref{alg:plug-and-play prior}. For $x$ steps, we use direct proximal operator in DIP-ADMM-v2, and ISTA in DIP-ADMM-v1. For $\theta$ step in DIP-ADMM-v1 (\ref{alg:DIP-ADMM-v1}) and DIP-ADMM-v2 (\ref{alg:DIP-ADMM-v2}), we actually use ADAM\cite{kingma2014adam} to take one descent step with learning rate 0.001. For $y$ step in plug-and-play prior (\ref{alg:plug-and-play prior}), since it is convex in the $y$ step of \ref{eq:admm_detail_3}, we can directly set the derivative to zero and get the closed form for minimization. For all $u$ steps, we take one gradient ascent step with learning rate 1, which is consistent with the setting in \cite{boyd2011distributed}. For the penalty coefficient $\rho$, we set it to 1.
\\
\\
We use CelebA dataset \cite{liu2015faceattributes} to test these algorithms. All the images are resized to $128\times 128 \times 3$ (49152 pixels). To get the measurement $b$, we first randomly choose half of the pixels, and add Gaussian noise with zero mean and standard deviation $\sigma=10$ (the image range is from 0 to 255). Therefore, the measurement matrix $A$ is a truncated permutation matrix, and the task can be regarded as image inpainting
\\
\\
In the experiment, we try BM3D\cite{dabov2007image}, NLM\cite{buades2005non}, and TV \cite{Barbero11,JMLR:v19:13-538} denoising algorithms as proximal operators. 10 images from CelebA \cite{liu2015faceattributes} are randomly chosen to tune the parameters in the proximal operator (or denoising algorithm). After tuning, we choose the shrinkage parameter 0.01 for TV proximal operator. For BM3D denoising algorithm, we choose noise standard deviation as 0.05. For NLM denoising algorithm, we choose noise standard deviation 0.01, patch distance 2, cut-off distance 0.05. For detailed description of these parameters, please refer to the implementations of these denoising algorithms at these web pages\footnote{BM3D:\url{http://www.cs.tut.fi/~foi/GCF-BM3D/}, NLM: skimage.restoration.denoise\_nl\_means at \url{https://scikit-image.org/docs/dev/api/skimage.restoration.html}, TV:\url{https://github.com/albarji/proxTV}}. We test DIP-ADMM-v1, DIP-ADMM-v2 on another randomly chosen 25 images from CelebA dataset \cite{liu2015faceattributes}. We use peak signal noise ratio (PSNR) as the metric for reconstructed images. PSNR is defined as
\begin{equation}
\text{PSNR} = 20 \cdot \log\Big( \frac{\text{max}(I_r)}{\sqrt{\text{mse}(I_r, I)}} \Big)
\end{equation}where $I$ is the reconstructed image and $I_r$ is the reference image. $\text{max}(I_r)$ is the maximum pixel values in the reference image $I_r$. $\text{mse}(I_r, I)$ is the mean squared error between the reference image $I_r$ and the reconstructed image $I$.
\\
\\
For the choice of the convolutional neural network, we use the similar encoder-decoder architecture to deep image prior \cite{ulyanov2018deep}. The parameters of the network are $n_u=n_d=[16, 32, 64, 128, 128], n_s=[0, 0, 0, 0, 0], k_d=k_u=3$. For the explanation of these parameters of the network, please refer to the deep image prior supplemental material\footnote{\url{http://openaccess.thecvf.com/content_cvpr_2018/Supplemental/2711-supp.pdf}}.
\begin{algorithm}
  \caption{DIP-ADMM-v1}
  \label{alg:DIP-ADMM-v1}
  \textbf{Input}:
  Augmented Lagrangian $\mathcal{L}_\rho^{(1)}$,
  Number of iterations $T$,
  Number of iterations in $x$ step $N$,
  Regularizer $R$,
  Convolutional Neural Network $G$,
  Primal initialization $\theta_0, x_0$,
  Dual initialization $u_0$,
  Step size for the weights $\beta$
  \begin{algorithmic}[1]
    \For{ $t=0,1,...,T-1$} 
    \State $x_{t+1} = $ ISTA($x_t$, $N$, $\mathcal{L}_\rho^{(1)}$, $R$)
    \State $\theta_{t+1} = \theta_t -\beta\nabla_{\theta}\mathcal{L}_\rho^{(1)}(x_{t+1}, \theta, u_t)|_{\theta=\theta_t}$
    \State $u_{t+1}=u_t + \big(x_{t+1} - G(\theta_{t+1})\big)$
    \EndFor
    \\
    \Function{ISTA}{$x_t$, $N$, $\mathcal{L}_\rho$, $R$}
    \For{ $i=0,1,...,N-1$} 
    \State $x_{t}^{(i+1)} = \text{prox}_{\frac{R}{\lVert A^TA\rVert+\rho}}
    \big(x_t^{(i)}-\frac{1}{\lVert A^TA\rVert+\rho}\nabla_x\mathcal{L}_{\rho}(x, \theta_t, u_t)|_{x=x_t^{(i)}}\big)$
    \EndFor
    \\
    \Return $x_t^{(n)}$
    \EndFunction
  \end{algorithmic}
\end{algorithm}

\begin{algorithm}
  \caption{DIP-ADMM-v2}
  \label{alg:DIP-ADMM-v2}
  \textbf{Input}:
  Augmented Lagrangian $\mathcal{L}_\rho^{(2)}$,
  Number of iterations $T$,
  Regularizer $R$,
  Convolutional Neural Network $G$,
  Primal initialization $\theta_0, x_0$,
  Dual initialization $u_0$,
  Step size for the weights $\beta$
  \begin{algorithmic}[1]
    \For{ $t=0,1,...,T-1$} 
    \State $x_{t+1}=\text{prox}_{\frac{R}{\rho}}(G(\theta_t)-u_t)$
    \State $\theta_{t+1} = \theta_t -\beta\nabla_{\theta}\mathcal{L}_\rho^{(2)}(x_{t+1}, \theta, u_t)|_{\theta=\theta_t}$
    \State $u_{t+1}=u_t + \big(x_{t+1} - G(\theta_{t+1})\big)$
    \EndFor
  \end{algorithmic}
\end{algorithm}

\begin{algorithm}
  \caption{plug-and-play prior}
  \label{alg:plug-and-play prior}
  \textbf{Input}:
  Augmented Lagrangian $\mathcal{L}_\rho^{(3)}$,
  Number of iterations $T$,
  Regularizer $R$,
  Primal initialization $y_0, x_0$,
  Dual initialization $u_0$,
  \begin{algorithmic}[1]
    \For{ $t=0,1,...,T-1$} 
    \State $x_{t+1}=\text{prox}_{\frac{R}{\rho}}(y_t-u_t)$
    \State $y_{t+1} = (A^TA+\rho I)^{-1}\big(A^Tb+\rho(x_{t+1}+u_t)\big)$
    \State $u_{t+1}=u_t + \big(x_{t+1} - y_{t+1}\big)$
    \EndFor
  \end{algorithmic}
\end{algorithm}

\section{Results}
\subsection{Comparison between DIP-ADMM-v1, DIP-ADMM-v2 and gradient descent}
\begin{figure}[htbp]
\minipage{0.33\textwidth}
  \includegraphics[width=\linewidth]{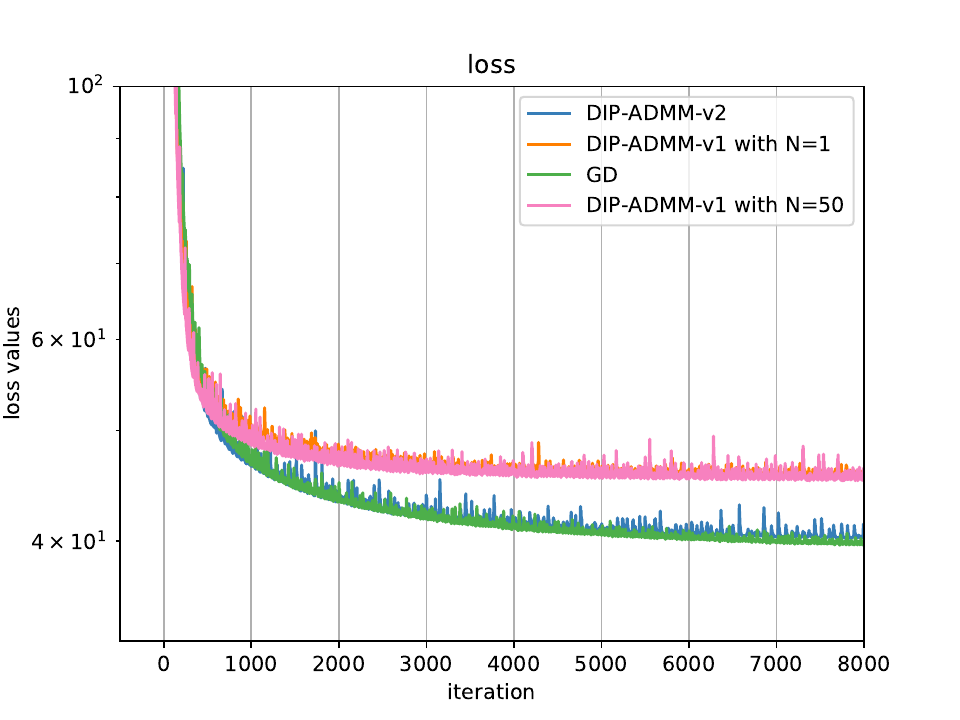}
\endminipage\hfill
\minipage{0.33\textwidth}
  \includegraphics[width=\linewidth]{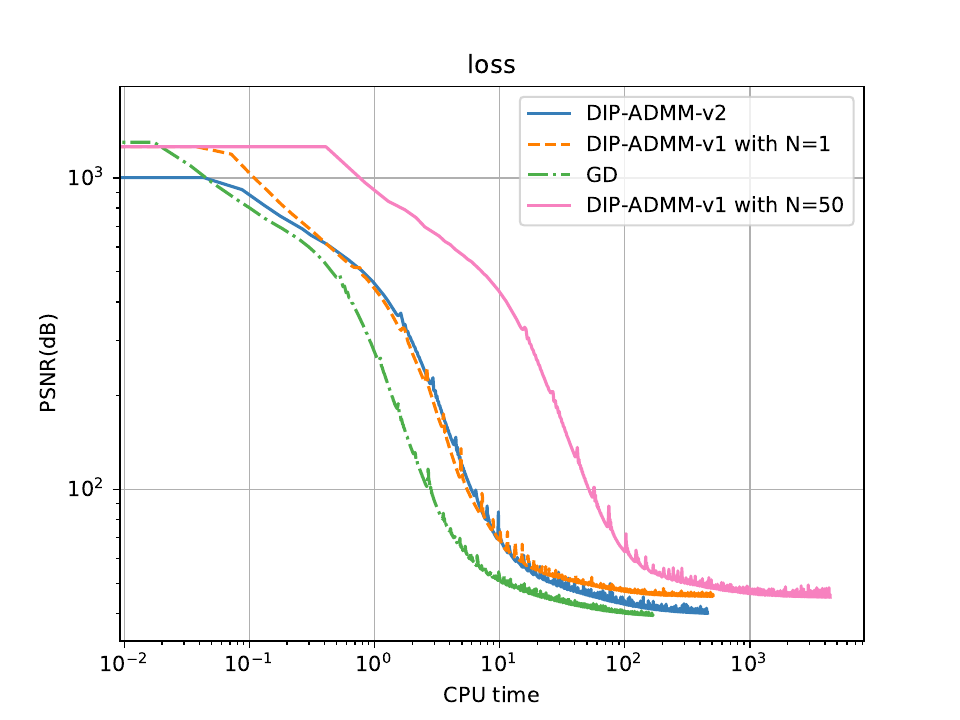}
\endminipage\hfill
\minipage{0.33\textwidth}
  \includegraphics[width=\linewidth]{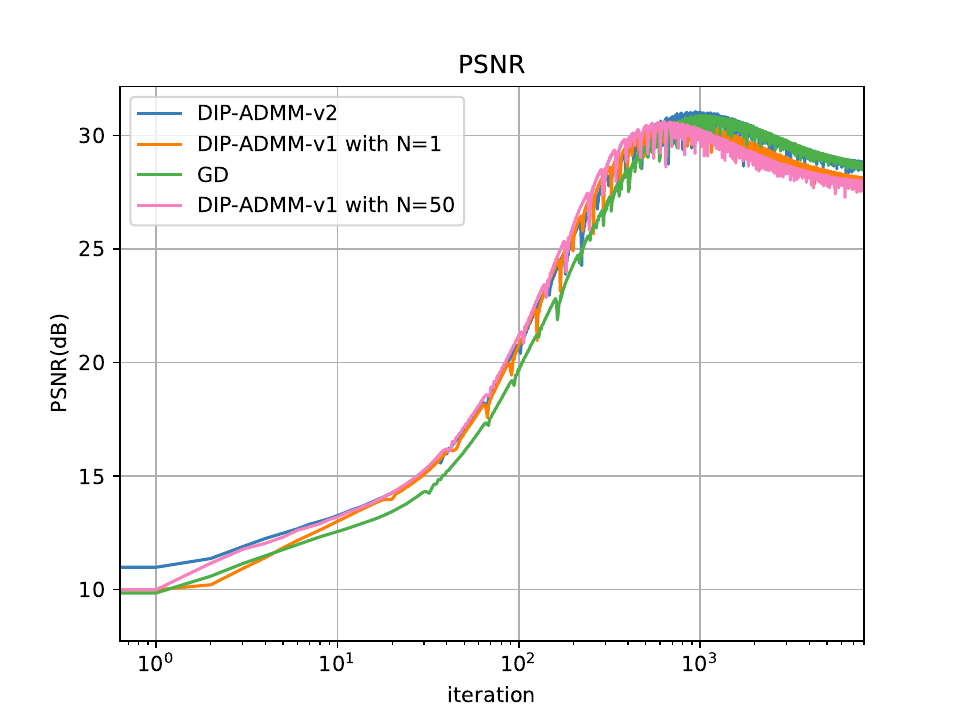}
\endminipage
  \caption{loss and PSNR for DIP-ADMM-v1, DIP-ADMM-v2, gradient descent}
  \label{fig:admm_compare}
\end{figure}

Both DIP-ADMM-v1 and DIP-ADMM-v2 can solve the problem for DIP with a prior, but DIP-ADMM-v1 needs more computation due to multiple evaluations of proximal operator while DIP-ADMM-v2 only needs to evaluate the proximal operator once per iteration. Figure \ref{fig:admm_compare} shows the numerical results to compare these two algorithms. We use TV prior to test these two algorithms and the loss functions is defined as 
\begin{equation}
L(\theta) = \frac{1}{2}\lVert b-AG(\theta)\rVert_2^2 + 0.01 \lVert G(\theta)\rVert_{TV}
\end{equation}
The number of ISTA iteration $N$ in the algorithm (\ref{alg:DIP-ADMM-v1}) does not have much influence. Regardless of the generalization, the ADMM-DIP-v2 has lower loss value, and the PSNR values for these algorithms are very close. ADMM-DIP-v2 and gradient descent have a very similar result, but the implementation is different. For gradient descent, it directly optimizes the loss function above with explicit TV prior. For ADMM-DIP-v2, it utilizes the existing TV denoising solver\footnote{Anisotropic Total Variation on a 2-dimensional signal (image denoising) in proxTV library at \url{https://github.com/albarji/proxTV}} as the proximal operator, and does not have any explicit TV prior formula. 
\\
\\
For DIP-ADMM-v1 and DIP-ADMM-v2, DIP-ADMM-v2 has better performance in both loss values and computation time. For DIP-ADMM-v2 and gradient descent, DIP-ADMM-v1 can incorporate some implicit priors such as denoising algorithm while gradient descent requires the explicit form of the prior, so DIP-ADMM-v2 can gives us more options to choose the desired priors.
\subsection{Comparison from learning curves}
\begin{figure}[!htb]
\minipage{0.5\textwidth}
  \includegraphics[width=\linewidth]{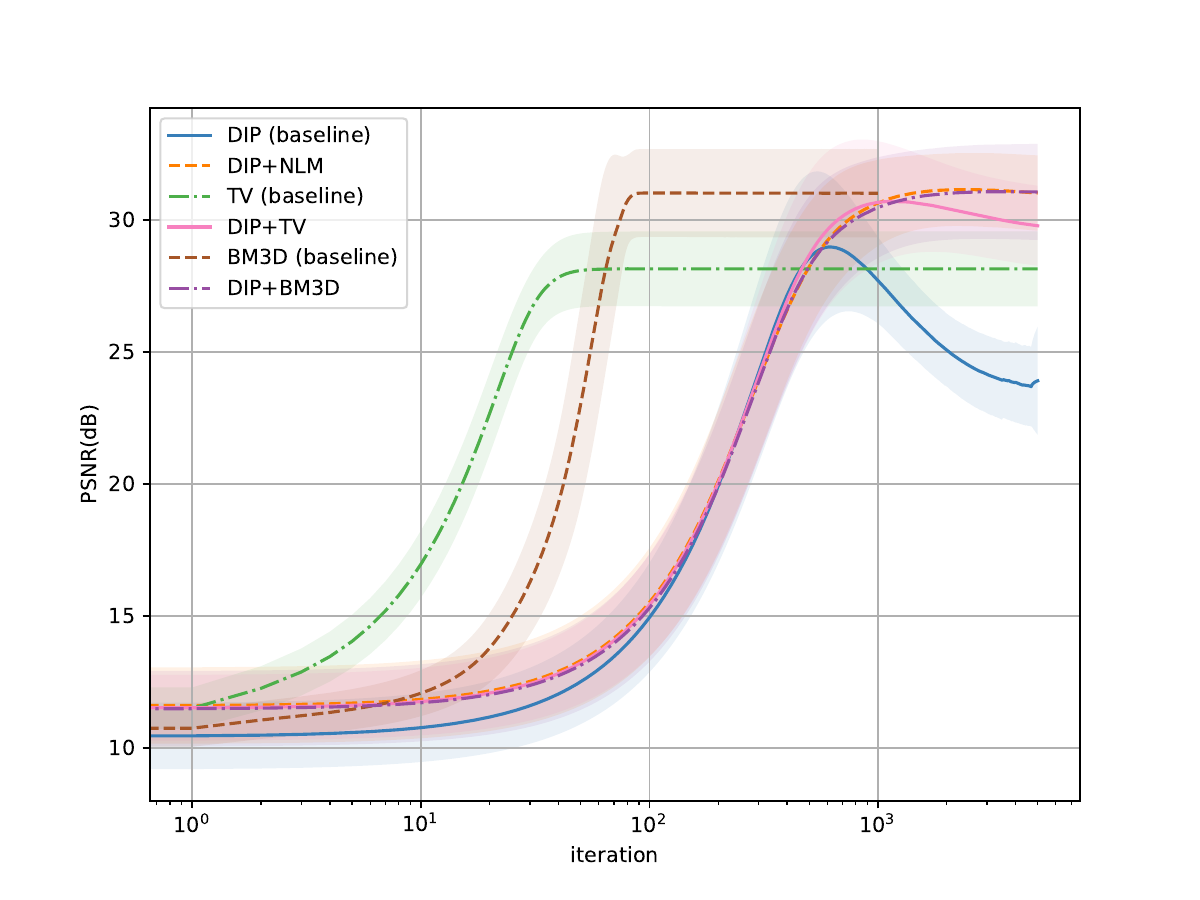}
  \caption{PSNR with respect to iteration}
  \label{fig:all_alg_compare_sub1}
\endminipage\hfill
\minipage{0.5\textwidth}
  \includegraphics[width=\linewidth]{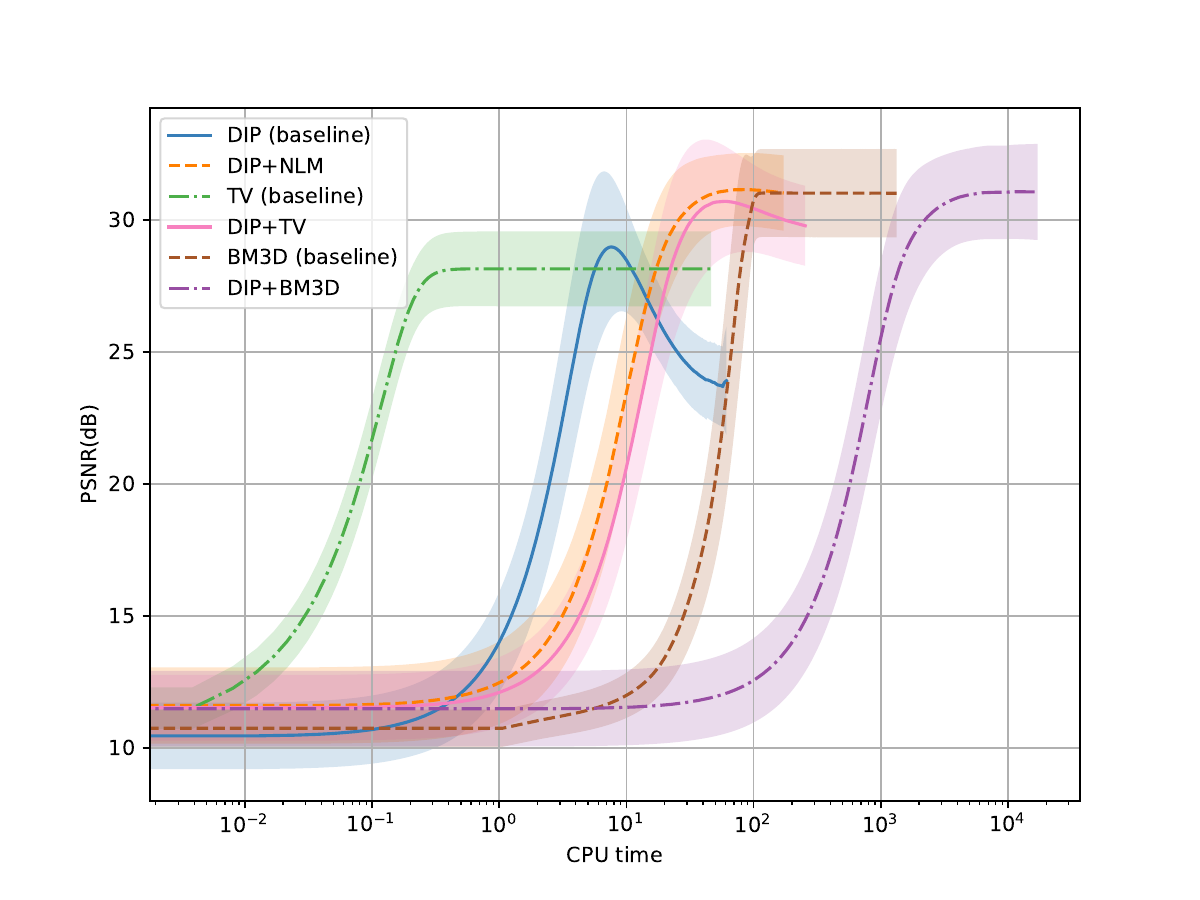}
  \caption{PSNR with respect to cpu time}
  \label{fig:all_alg_compare_sub2}
\endminipage
\end{figure}

In the following result parts, we will only use DIP-ADMM-v2 to incorporate the DIP with a prior. Also, we will consider to reconstruct the image only with a denoising algorithm such as TV by plug-and-play prior algorithm \cite{venkatakrishnan2013plug} (cf. algorithm \ref{alg:plug-and-play prior}). As a baseline, we use gradient descent to reconstruct images with original deep image prior. We test these algorithms on 25 images chosen from CelebA dataset\cite{liu2015faceattributes}. Figure \ref{fig:all_alg_compare_sub1} and \ref{fig:all_alg_compare_sub2} show the PSNR with respect to the iteration and CPU time. The curves are averaged among these 25 images to get the mean and standard deviation (shade in the figures). For DIP+denoising algorithm, we use DIP-ADMM-v2 (\ref{alg:DIP-ADMM-v2}). For TV and BM3D \cite{dabov2007image}, we use plug-and-play prior algorithm (\ref{alg:plug-and-play prior}). Note that we don't plot the curve for NLM denoising algorithm\cite{buades2005non} because it fails to reconstruct the image by plug-and-play prior algorithm(\ref{alg:plug-and-play prior}), but it can work with deep image prior by DIP-ADMM-v2 (\ref{alg:DIP-ADMM-v2}). we also use $r_{t+1}=0.9r_t + 0.1x_t$  to smooth the result $x_t$ of each iteration so the PSNR curves look smooth.
\\
\\
From figures \ref{fig:all_alg_compare_sub1} and \ref{fig:all_alg_compare_sub2}, DIP will overfit after several iterations. If we combine DIP with some denoising algorithms, the PSNR curves will be almost flat when it reaches the peak. We can also use denoising algorithm alone. If we use BM3D \cite{dabov2007image} via plug-and-play prior (\ref{alg:plug-and-play prior}) for the inpainting task, the performance is superior to other methods except DIP+BM3D. However, the gap between DIP+BM3D and BM3D is very small, and the final PSNR for DIP+BM3D is only 0.05 dB larger then BM3D. When DIP is combined BM3D, the computation time will increase largely. Maybe NLM \cite{buades2005non} and TV are better choices since they do not add an excessive computational burden. BM3D can also utilize parallel computation for acceleration \cite{honzatko2019accelerating}, which means the computation time caused by denoising algorithms could be reduced.
\subsection{Comparison from images}
\begin{figure}[htbp]
\centering
  \includegraphics[width=0.5\linewidth]{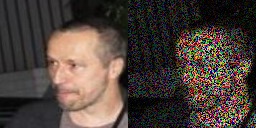}
  \caption{true image and corrupted image with missing pixels}
  \label{fig:true_corrupted_img}
\end{figure}

\begin{figure}[htbp]
\centering
\includegraphics[width=\linewidth]{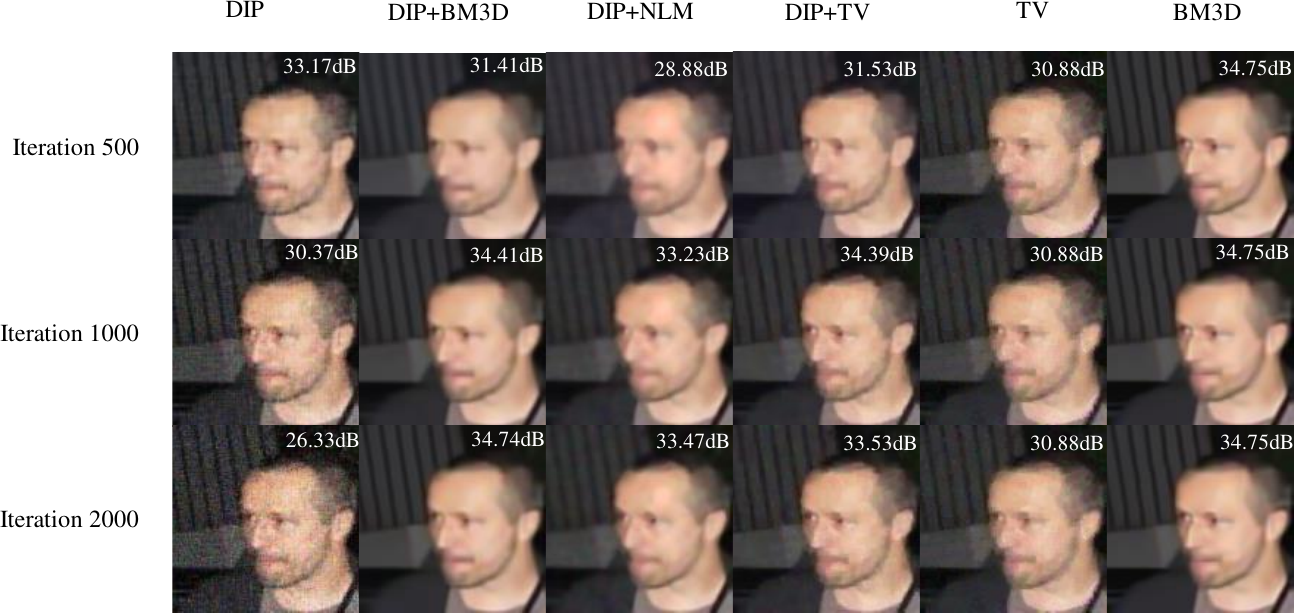}
\caption{image evolution for these algorithms}
\label{fig:celeba_img_result}
\end{figure}

\begin{figure}[htbp]
\centering
  \includegraphics[width=0.5\linewidth]{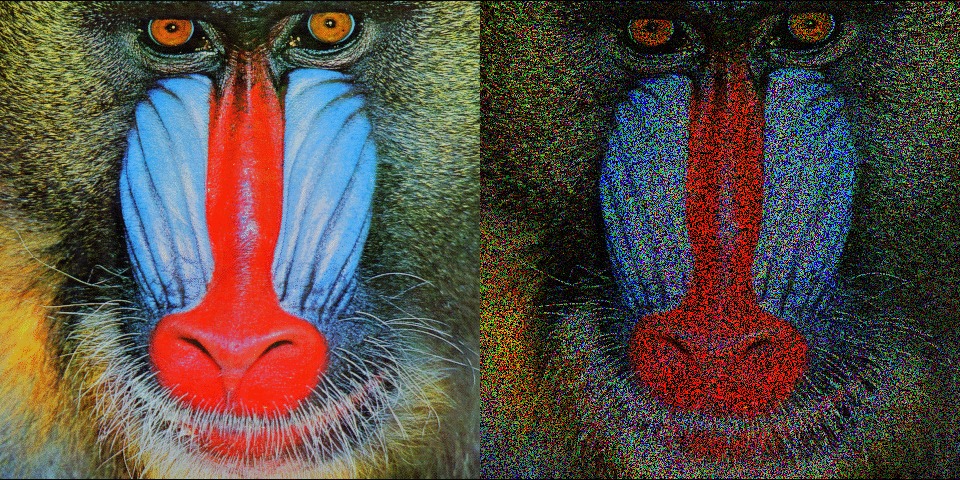}
  \caption{true image and corrupted image with missing pixels}
  \label{fig:true_corrupted_img_hi_res}
\end{figure}

\begin{figure}[!htb]
\minipage{0.16\textwidth}
  \includegraphics[width=\linewidth]{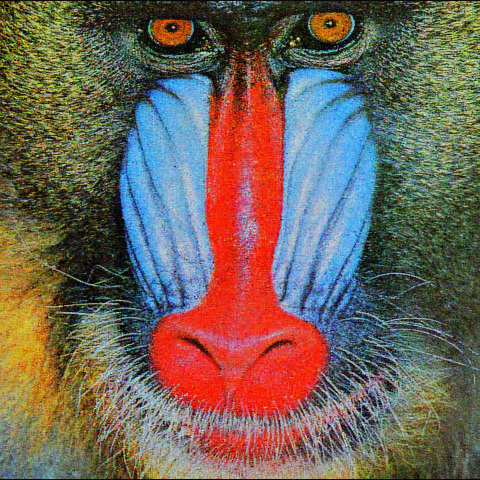}
\endminipage\hfill
\minipage{0.16\textwidth}
  \includegraphics[width=\linewidth]{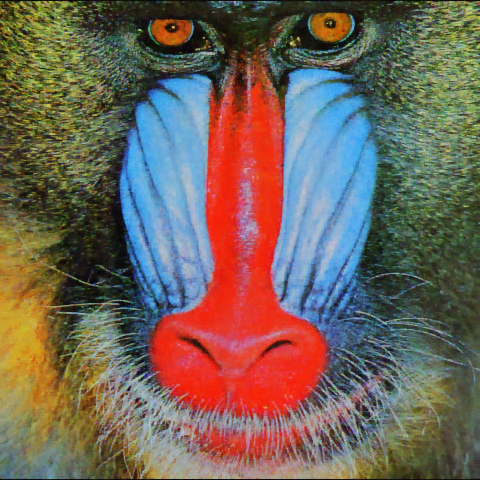}
\endminipage\hfill
\minipage{0.16\textwidth}
  \includegraphics[width=\linewidth]{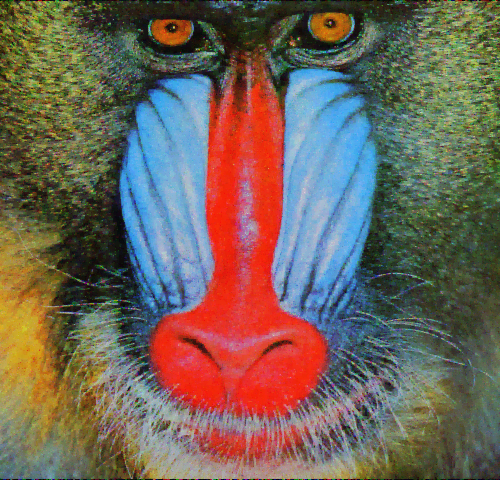}
\endminipage\hfill
\minipage{0.16\textwidth}
  \includegraphics[width=\linewidth]{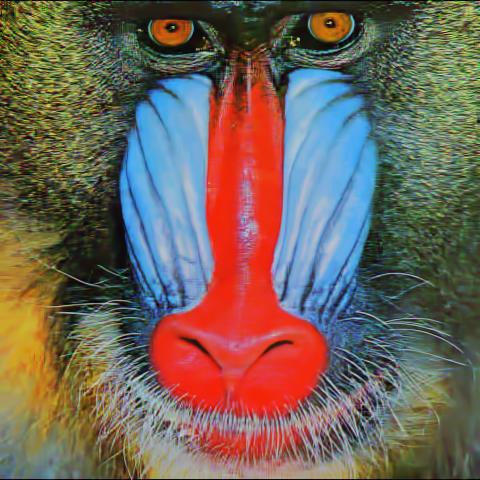}
\endminipage\hfill
\minipage{0.16\textwidth}
  \includegraphics[width=\linewidth]{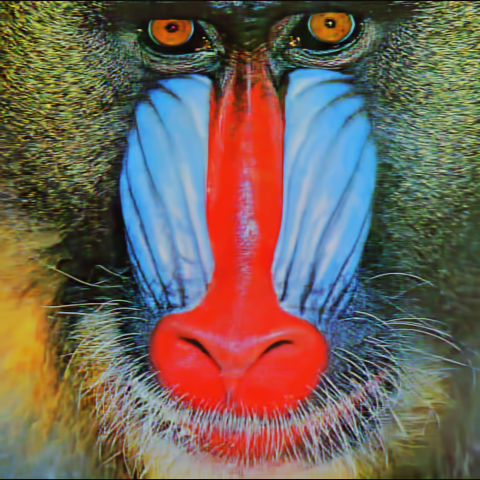}
\endminipage\hfill
\minipage{0.16\textwidth}
  \includegraphics[width=\linewidth]{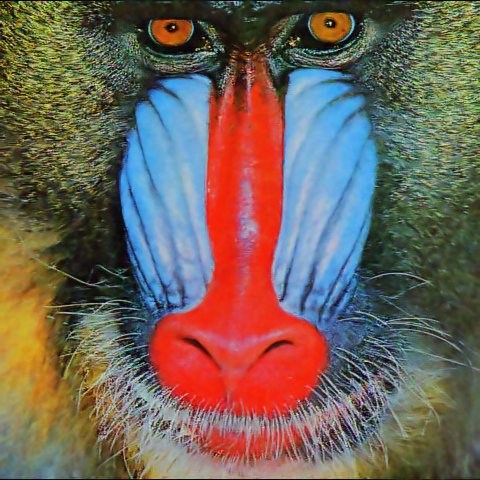}
\endminipage
\caption{Results(from left to right) for DIP(22.15dB), DIP+TV(23.91dB), TV(22.64dB), BM3D(23.46dB), DIP+BM3D(23.73dB), DIP+NLM(23.97dB)}
\label{fig:hi-res-compare}
\end{figure}

Figures \ref{fig:true_corrupted_img} show the true image and the corrupted image with 50\% missing pixels. Note that the corrupted image is not the measurement $b$. It can only be regarded as a zero-fill reconstruction, and the measurement $b$  does not have these zeros filled in the corrupted image. Figure \ref{fig:celeba_img_result} shows the image evolution at iteration 500, 1000, 2000 for these prior combinations. For only DIP, the image becomes noisy for larger iterations. If we combine DIP with other priors, the image can avoid overfitting. Even though you do not have early stopping, the PSNR can still be high. Note that the PSNR for TV and BM3D is static in Figure \ref{fig:celeba_img_result} because they have reached a plateau before iteration 500.
\\
\\
We also test the results for higher resolution images with shape $480 \times 480 \times 3$. The image evolution is similar to the Figure \ref{fig:celeba_img_result}, so briefly we will only show the result at iteration 5000 in Figure \ref{fig:hi-res-compare}. The true and corrupted image in Figure \ref{fig:true_corrupted_img_hi_res}. To see the difference of these results, please zoom in and you can observe that DIP has overfitted. As you can observe in Figure \ref{fig:all_alg_compare_sub2}, BM3D\cite{dabov2007image} is very slow, so for higher resolution image, it is not suitable to use BM3D as the proximal operator.


\begin{figure}[!htb]
\minipage{0.16\textwidth}
  \includegraphics[width=\linewidth]{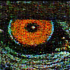}
\endminipage\hfill
\minipage{0.16\textwidth}
  \includegraphics[width=\linewidth]{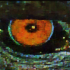}
\endminipage\hfill
\minipage{0.16\textwidth}
  \includegraphics[width=\linewidth]{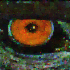}
\endminipage\hfill
\minipage{0.16\textwidth}
  \includegraphics[width=\linewidth]{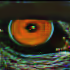}
\endminipage\hfill
\minipage{0.16\textwidth}
  \includegraphics[width=\linewidth]{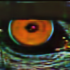}
\endminipage\hfill
\minipage{0.16\textwidth}
  \includegraphics[width=\linewidth]{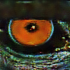}
\endminipage
\caption{Zoom-in (from left to right) for DIP, DIP+TV, TV, BM3D, DIP+BM3D, DIP+NLM}
\end{figure}
\section{DIP overfitting analysis}
In this section, the theoretical results are from \cite{heckel2019denoising}. We present the numerical results for images while \cite{heckel2019denoising} gives some of the result for 1D signals. For more details, please refer to the paper \cite{heckel2019denoising}. All the analysis is for denoising, which means the measurement matrix $A$ is an identity matrix. We then also use our results for the interpretation of overfitting and try to fit along some directions to avoid overfitting.

\subsection{Dynamics of deep image prior}
For the simplicity of analysis and better visualization of the results, we only consider the denoising problem, which means the measurement matrix $A$ is an identity matrix. The loss function is
\begin{equation}
L(\theta)=\frac{1}{2}\lVert G(\theta)-b\rVert_2^2
\end{equation}
The noisy image is  $b \in \mathbb{R}^n$, and the weights are $\theta \in \mathbb{R}^w$. The network $G(\theta)$ is over-parameterized\footnote{the neural network we used has 980787 parameters and the number of pixels for $128 \times 128 \times 3$ image is 49152. The number of parameters is much larger than the number of pixels.}, which means $w\gg n$. Some papers \cite{jacot2018neural,heckel2019denoising} propose that over-parameterized networks can be well approximated by the corresponding linearized network, so we can rewrite our loss function with the linearized network.
\begin{equation}
L(\theta)_{lin}=\frac{1}{2}\lVert G(\theta_0)+J(\theta_0)(\theta-\theta_0)-b\rVert^2_2
\end{equation}$\theta_0$ is the random initialization of the network parameters. $J(\theta)$ is the Jacobian of the network $G(\theta)$ with respect to $\theta$, which is defined as $[J(\theta)]_{i,j}=\frac{\partial G(\theta)_i}{\partial \theta_j}$. Since the Jacobian is constant with random initialized weights for over-parameterized networks \cite{jacot2018neural,heckel2019denoising}, in the following part, for simplicity, we use $\mathbf{J}$ to replace $J(\theta_0)$ as the Jacobians. If we do gradient descent with the learning rate $\eta$, we can get the residuals with respect to the iteration. From \cite{heckel2019denoising}, the residuals $r_t, \hat{r}_t$ for the original loss and the linearized loss can be 
\begin{equation}
\begin{split}
r_t = G(\theta_t) - b
\\
\hat{r}_t = (I - \eta \mathbf{J}\mathbf{J}^T)^t r_0
\end{split}
\end{equation}
Since we approximate $G(\theta)$ with a linear model, we can also use $\hat{r}_t$ to approximate $r_t$. If we do singular value decomposition on $\mathbf{J}=U\Sigma V^T$, so $\mathbf{J}\mathbf{J}^T=U\Sigma^2 U^T$. We can rewrite the residual as
\begin{equation}
r_t \approx U(I-\eta\Sigma^2)^tU^Tr_0
\end{equation}
If $u_i$ for $i=1,2,...,n$ are the singular vectors in $U$, the residual can be represented as
\begin{equation}
r_t \approx \sum_{i=1}^{n}(1-\eta\sigma^2_i)^t \langle u_i, r_0 \rangle u_i
\label{eq:final_residual}
\end{equation}From this residual, if the learning rate $0 < \eta \leq \frac{1}{\sigma^2_{max}}$  and $t\rightarrow \infty$, $r_t \rightarrow 0$. Therefore, $G(\theta_\infty)$=b, which means the neural network will fit the noisy image perfectly and the overfitting definitely happens when t is infinite. In the numerical experiment, we also observe overfitting happens when the number of iteration is large.
\subsection{Generalization error with assumptions}
If the noisy image $b \in \mathbb{R}^n$ is obtained by $b=x^\natural + \epsilon$ where $\epsilon$ is the noise with Gaussian distribution $\mathcal{N}(\mathbf{0}, \zeta^2 \mathbf{I})$. The generalization error for DIP is
\begin{equation}
\begin{split}
\lVert G(\theta_t)-x^{\natural}\rVert_2&=\lVert G(\theta_t)-b+\epsilon \rVert_2
\\
&=\lVert r_t+\epsilon\rVert_2
\\
&=\lVert U(I-\eta\Sigma^2)^tU^Tr_0 + \epsilon\rVert_2
\\
&=\lVert U(I-\eta\Sigma^2)^tU^T(G(\theta_0)-x^\natural-\epsilon) + \epsilon\rVert_2
\\
&=\lVert U(I-\eta\Sigma^2)^tU^TG(\theta_0) - U(I-\eta\Sigma^2)^tU^Tx^\natural + U[I-(I-\eta\Sigma^2)^t]U^T \epsilon\rVert_2
\\
&\leq \lVert U(I-\eta\Sigma^2)^tU^TG(\theta_0) \rVert_2 + \lVert U(I-\eta\Sigma^2)^tU^Tx^\natural \rVert_2 + \lVert U[I-(I-\eta\Sigma^2)^t]U^T \epsilon\rVert_2
\end{split}
\end{equation}
If we have the assumption that $x^\natural$ can be spanned by only the singular vectors $u_i$ corresponding to the first $p$ leading singular values, which means $x^\natural = \overline{U}\overline{U}^Tx^\natural$, where $\overline{U}=[u_1, u_2, ..., u_p]$. We will prove this assumption is reasonable in the experiment results. The generalization error can then be
\begin{equation}
\begin{split}
\lVert G(\theta_t)-x^{\natural}\rVert_2 
\leq 
(1-\eta \sigma_{min}^2)^t\lVert G(\theta_0)\rVert_2
+ (1-\eta \sigma_{p}^2)^t\lVert x^\natural\rVert_2
+\sqrt{\sum_{i=1}^{n}[1-(1-\eta \sigma_i^2)^t]^2\langle u_i, \epsilon \rangle^2}
\end{split}
\end{equation}The first term on the right hand side is only related to the network initialization $\theta_0$. The second term is related the the true image and the third term is related the noise. Comparing the second and third term, we can see the second term will decay as the number of iteration increases while the third term will increase. However, for some different singular values, the speed of increasing in the third term is different. When the second term is almost zero, the third term becomes large and the network is still fitting noise, so one can observe that the network will first mostly fit the true image and then the noise. This generalization bound and explanation is also at \cite{heckel2019denoising}
\subsection{Singular value decomposition of the Jacobian}
\begin{figure}[htbp]
\centering
  \includegraphics[width=0.5\linewidth]{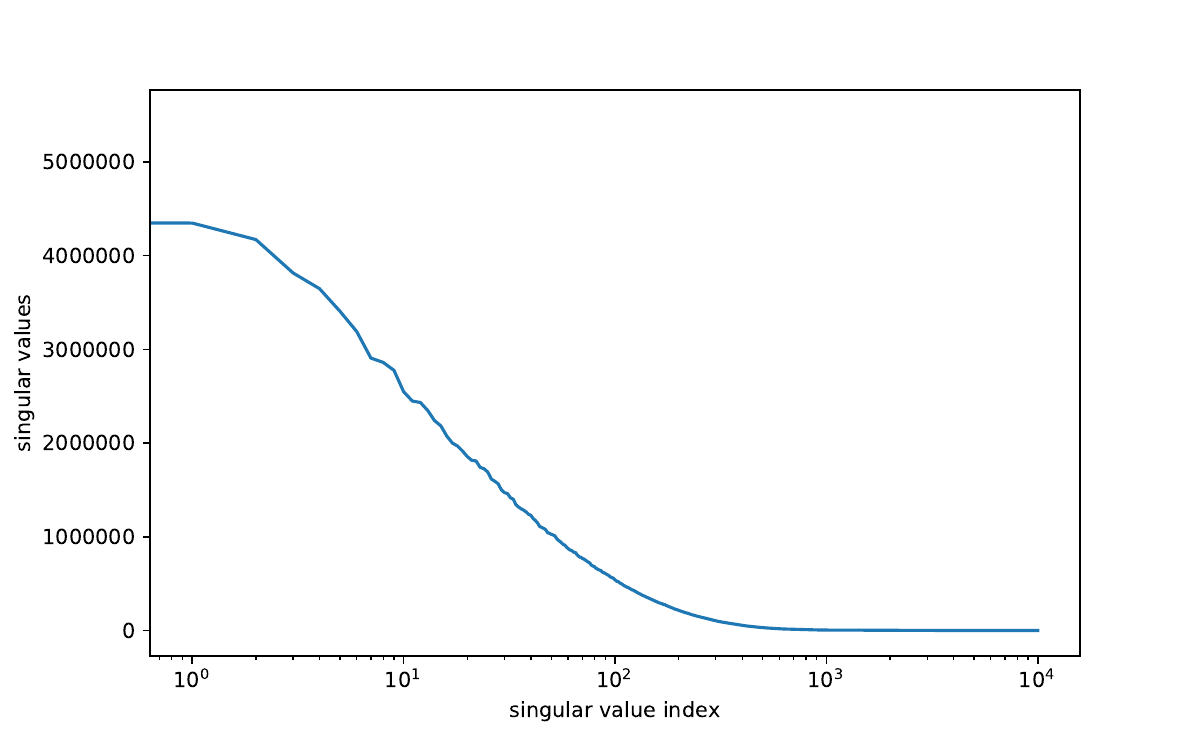}
  \caption{Singular values}
  \label{fig:sv}
\end{figure}

\begin{figure}[!htb]
\minipage{0.2\textwidth}
  \includegraphics[width=\linewidth]{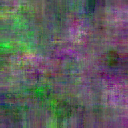}
\endminipage\hfill
\minipage{0.2\textwidth}
  \includegraphics[width=\linewidth]{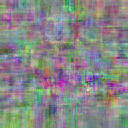}
\endminipage\hfill
\minipage{0.2\textwidth}
  \includegraphics[width=\linewidth]{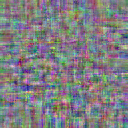}
\endminipage\hfill
\minipage{0.2\textwidth}
  \includegraphics[width=\linewidth]{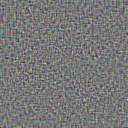}
\endminipage
\caption{The 10th, 100th, 1000th, and 10000th singular vectors}
\label{fig:s_vector}
\end{figure}

We do the singular value decomposition of $\mathbf{J}\mathbf{J}^T$ and get the first 10000 largest singular values and corresponding singular vectors. For the implementation details, we first need to get $\mathbf{J}\mathbf{J}^T$. Actually, getting the explicit form of $\mathbf{J}\mathbf{J}^T$ is computational expensive, so we only regard $\mathbf{J}\mathbf{J}^T$ as linear operator and only get the function which can calculate $\mathbf{J}\mathbf{J}^Tx$. In Pytorch, we can use autograd backward function to get vector-Jacobian product $\mathbf{J}^Ty$ and Jacobian-vector product\footnote{It is not direct to get $\mathbf{J}x$ since it needs to use backward twice. For the first time, we get $\mathbf{J}^Ty$. For the second time, we use backward to get $\frac{\partial\mathbf{J}^Ty}{\partial y}|_{y=x}=\mathbf{J}x$} $\mathbf{J}x$. To get the operator of $\mathbf{J}\mathbf{J}^T$, we combine vector-Jacobian and Jacobian-vector products. To get the first 10000 largest singular values and vectors, we use ARPACK\cite{lehoucq1998arpack}, which only needs this linear operator. 
\\
\\
Figure \ref{fig:sv} and \ref{fig:s_vector} shows the first 10000 singular values and vectors\footnote{The vectors are reshaped to $128 \times 128 \times 3$ for image visualization}. Obviously, $\mathbf{J}\mathbf{J}^T$ is low rank since the singular values have a cut-off. The singular vectors have smooth representation at larger singular values. As the singular value decreases, the singular vectors have more high frequency components. 

\subsection{Different representations for image and noise}
\begin{figure}[!htb]
\minipage{0.4\textwidth}
  \includegraphics[width=\linewidth]{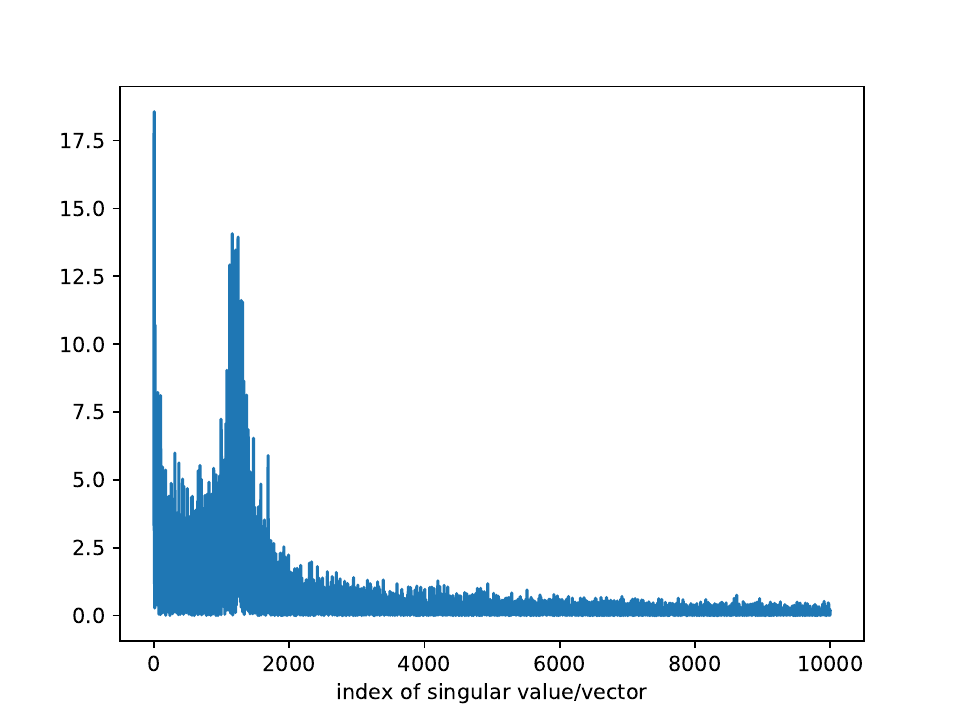}
\endminipage\hfill
\minipage{0.4\textwidth}
  \includegraphics[width=\linewidth]{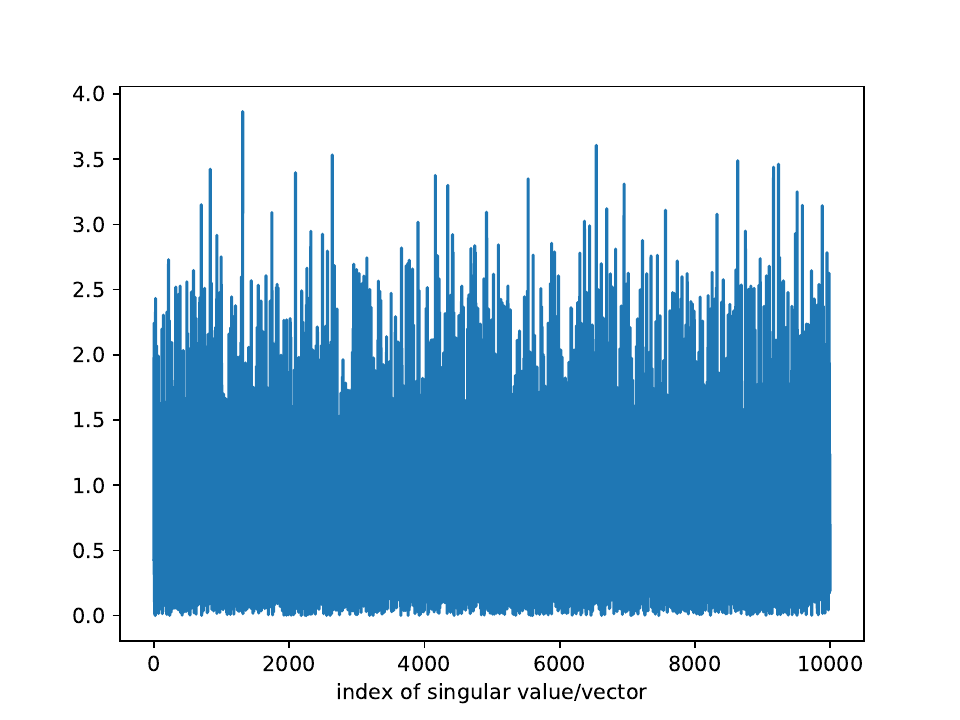}
\endminipage
\caption{Projection on the space spanned by singular vectors $U$ for image(left) and noise(right)}
\label{fig:projection_img_noise}
\end{figure}

We will use some numerical results to show the assumption that $x^\natural$ can be spanned by only the singular vectors $u_i$ corresponding to the first $p$ leading singular values is reasonable. Also, the noise should be spanned by almost all singular vectors. To test this assumption, we should first get the results of $U^Tx^\natural$ and $U^T\epsilon$. In the last subsection, we get the part of matrix $U$ including first 10000 singular vectors, so we can get the first 10000 elements of vector $U^Tx^\natural$ and $U^T\epsilon$. Figure \ref{fig:projection_img_noise} shows the projection result. The image has large values at the positions where the singular values are large while the noise is evenly spreading in the space. The numerical results can only show the assumption is approximately reasonable since the image should also have extra components beyond the $p$ leading singular vectors.
The results (\ref{eq:final_residual}) in the last subsection explains that deep image prior will converge faster at the direction where the singular values are large, which means most components at lower indices will be fitted and the residual at these directions will approach zero first. As the number of iterations increases, components at larger indices will also be fitted. As the number of iterations goes to infinity, all the components will be fitted and the residuals at all directions will be zero. The image has large components at the lower indices, which means the image will be fitted first, while the noise has components evenly spreading in the space and will be fitted partly. However, as the number of iteration increases, the model will mainly fit the noise since most of the components in image has been fitted.
\\
\\
For the interpretation of the behavior of deep image prior, it works like a dynamic low-pass filter. Normal low-pass filter is performed in the Fourier domain and will not change with time. However, this deep image prior is in an unknown domain, which, however, has some similarities with Fourier domain. On the other side, the low-pass filter from DIP is not static and the bandwidth will become large as the number of iteration increases. As the number of iteration goes to infinity, the bandwidth covers all the components in the domain, which is an all-pass filter. In this case, the result from DIP will be exactly the noisy image.

\subsection{Fitting along some directions}
\begin{figure}[!htb]
\minipage{0.4\textwidth}
  \includegraphics[width=\linewidth]{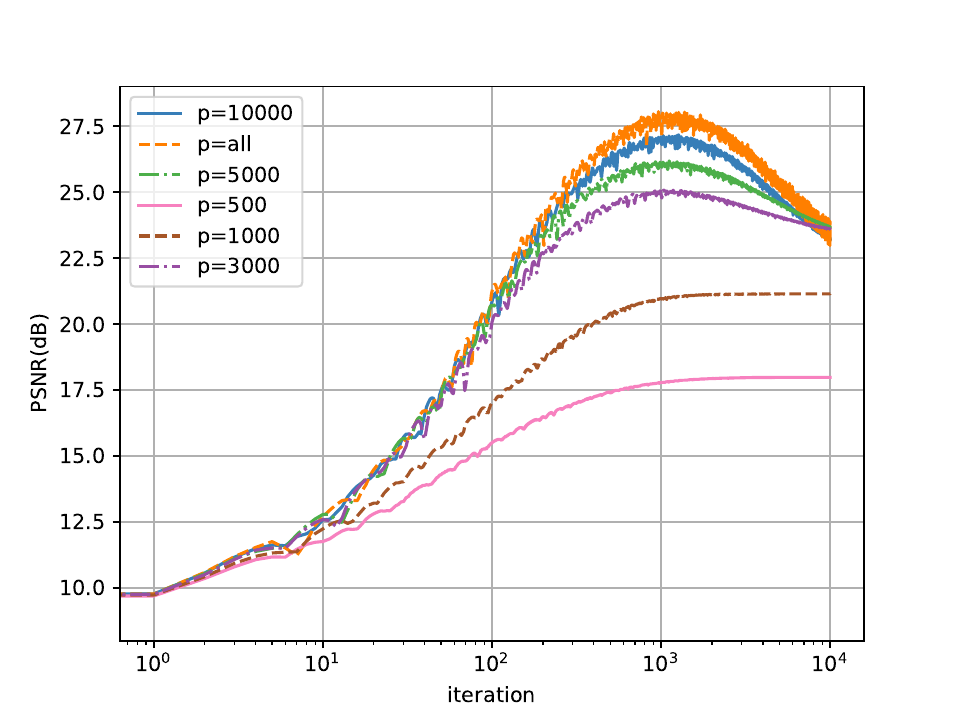}
\endminipage\hfill
\minipage{0.4\textwidth}
  \includegraphics[width=\linewidth]{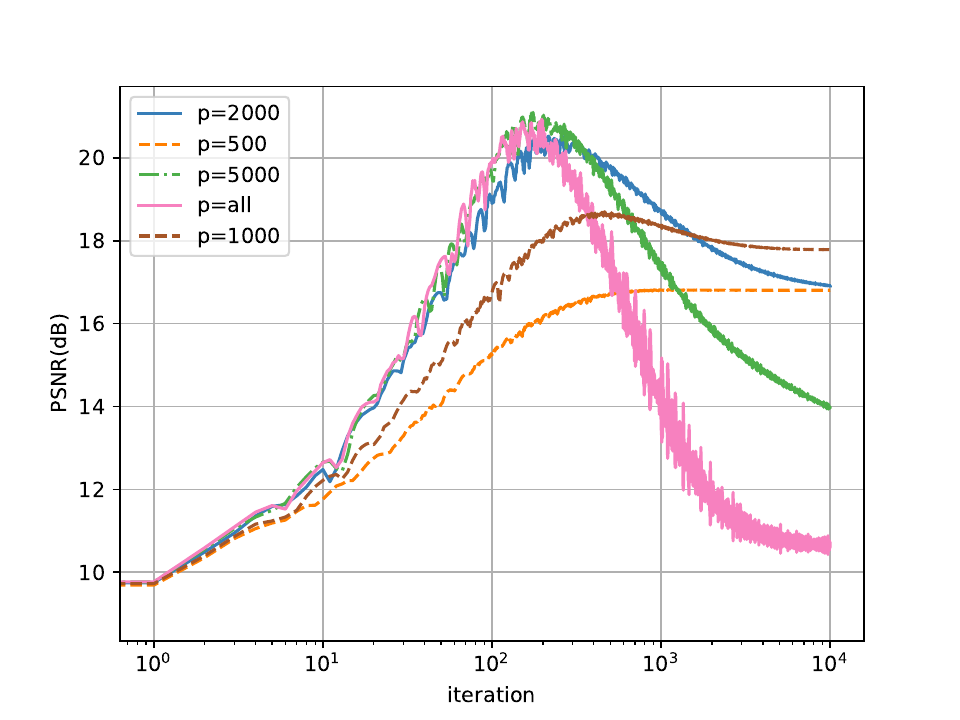}
\endminipage
\caption{PSNR curves for noise standard deviation 25(left) and 100(right)}
\label{fig:direction_loss}
\end{figure}

If we can only fit the image in some directions corresponding to the several largest singular values, it might prevent overfitting. In this section, we will use deep image prior for denoising in some directions. The loss function is defined below.
\begin{equation}
L(\theta) = \frac{1}{2}\lVert \overline{U}^T(G(\theta)-b)\rVert_2^2
\end{equation}
 $\overline{U} \in \mathbb{R}^{n\times p}$ only has the first $p$ singular vectors $[u_1, u_2, ..., u_p]$, which make sure $G(\theta)$ only fits noisy image $b$ at the specified directions. Figure \ref{fig:direction_loss} shows the PSNR curves for different noise levels and singular vectors. When the noise level is large, this method has some effect on avoiding overfitting. When the noise level is small, the effect is not obvious. When $p$ is large, the effect of preventing overfitting is weakened. When p is small, the reconstruction result will become worse.
 \\
 \\
 We can imagine two limit cases. When there is no noise, we can directly fit the image with $U^T$. Since this is an orthogonal matrix, the loss function will be the original one $\frac{1}{2}\lVert G(\theta)-b\rVert_2^2$  and the solution will be exactly the original image without noise. If we only fit along some directions, this may be detrimental to the performance. When the noise level is infinite, the original image is completely overwhelmed by noise so the measurement $b$ can be regarded as pure noise. For this case, there is no need to fit the noise and $\overline{U}^T$can be a zero matrix. If the noise is large, the image at the directions of the first singular vectors is not totally overwhelmed while the image projected to other directions is overwhelmed. For this case, it is necessary to fit the image along the the first singular vectors.
\section{Conclusion}
We mainly study the overfitting of deep image prior and find some solutions to tackle it. We propose two algorithms called DIP-ADMM-v1 and DIP-ADMM-v2 to incoporate an implicit or explicit prior with deep image prior for preventing overfitting. Compared with DIP-ADMM-v1, DIP-ADMM-v2 can exactly minimize one step in ADMM algorithm and also has better performance in both reconstruction and computation time. These algorithms can combine implicit priors such as denoising algorithms, which means they have more flexibility and options to choose priors. From experiment results, when DIP is combined with a denoising algorithm as a prior, the overfitting effect is mitigated or almost disappears. The computation cost caused by denoising algorithms is acceptable when we choose proper denoising algorithms or use efficient algorithm implementation.
\\
\\
We also analyze the overfitting of DIP from gradient descent dynamics. The Jacobian of deep image prior plays an important role in this analysis. The over-parameterization of deep image prior makes it possible to linearize DIP, and the Jacobian can be regarded as constant. The singular values of Jacobian determines the fitting speed along the corresponding singular vectors. Images and noise have different representations in the space spanned by singular vectors so that we can do optimization only along some directions to avoid overfitting. When the noise level is large, this kind of fitting is effective.
\\
\\
For the extension of the thesis, there are two points. The first one is that one could also incorporate a learned prior by a learned denoiser, which gives more control on the prior since one can determine the dataset to train the denoiser. On the other side, one can also use a pretrained denoiser even though one doesn't have enough data to train a denoiser. The learned prior is also proposed by \cite{van2018compressed}, but they use Tikhonov regularization with learned covariances and means to regularize the weights of DIP. It is nontrivial to get the such a learned prior. By incorporating a learned denoiser as a learned prior, it is more direct and convenient than \cite{van2018compressed}. 
\\
\\
Another point is that one can also use DIP-ADMM algorithms for neural network training with a large dataset. Actually, the common regularization methods are mainly performed on weights of the deep network such as L2 regularization. The weights of a network is abstract, so it is difficult to design a good prior. However, DIP-ADMM regularizes the output of a network, and the output is usually image, signal, or words, which are easier and intuitive for someone to design a prior such as denoising algorithms. Also, when we have a constraint on the output of the neural network, the proximal operator can be a projection onto the feasible domain of the neural network output.
\\
\\
\printbibliography

@inproceedings{gomez2019fast,
  title={Fast and Provable ADMM for learning with Generative Priors},
  author={Latorre, Fabian and Cevher, Volkan and others},
  booktitle={Advances in Neural Information Processing Systems},
  pages={12004--12016},
  year={2019}
}

@inproceedings{ulyanov2018deep,
  title={Deep image prior},
  author={Ulyanov, Dmitry and Vedaldi, Andrea and Lempitsky, Victor},
  booktitle={Proceedings of the IEEE Conference on Computer Vision and Pattern Recognition},
  pages={9446--9454},
  year={2018}
}

@article{metzler2018unsupervised,
  title={Unsupervised Learning with Stein's Unbiased Risk Estimator},
  author={Metzler, Christopher A and Mousavi, Ali and Heckel, Reinhard and Baraniuk, Richard G},
  journal={arXiv preprint arXiv:1805.10531},
  year={2018}
}

@article{kingma2014adam,
  title={Adam: A method for stochastic optimization},
  author={Kingma, Diederik P and Ba, Jimmy},
  journal={arXiv preprint arXiv:1412.6980},
  year={2014}
}

@inproceedings{venkatakrishnan2013plug,
  title={Plug-and-play priors for model based reconstruction},
  author={Venkatakrishnan, Singanallur V and Bouman, Charles A and Wohlberg, Brendt},
  booktitle={2013 IEEE Global Conference on Signal and Information Processing},
  pages={945--948},
  year={2013},
  organization={IEEE}
}

@article{dabov2007image,
  title={Image denoising by sparse 3-D transform-domain collaborative filtering},
  author={Dabov, Kostadin and Foi, Alessandro and Katkovnik, Vladimir and Egiazarian, Karen},
  journal={IEEE Transactions on image processing},
  volume={16},
  number={8},
  pages={2080--2095},
  year={2007},
  publisher={IEEE}
}

@article{boyd2011distributed,
  title={Distributed optimization and statistical learning via the alternating direction method of multipliers},
  author={Boyd, Stephen and Parikh, Neal and Chu, Eric and Peleato, Borja and Eckstein, Jonathan},
  journal={Foundations and Trends{\textregistered} in Machine learning},
  volume={3},
  number={1},
  pages={1--122},
  year={2011},
  publisher={Now Publishers Inc.}
}

@inproceedings{liu2015faceattributes,
 title = {Deep Learning Face Attributes in the Wild},
 author = {Liu, Ziwei and Luo, Ping and Wang, Xiaogang and Tang, Xiaoou},
 booktitle = {Proceedings of International Conference on Computer Vision (ICCV)},
 month = {December},
 year = {2015} 
}

@inproceedings{buades2005non,
  title={A non-local algorithm for image denoising},
  author={Buades, Antoni and Coll, Bartomeu and Morel, J-M},
  booktitle={2005 IEEE Computer Society Conference on Computer Vision and Pattern Recognition (CVPR'05)},
  volume={2},
  pages={60--65},
  year={2005},
  organization={IEEE}
}

@inproceedings{Barbero11,
  author = {Barbero, \'Alvaro and Sra, Suvrit},
  booktitle = {ICML},
  pages = {313-320},
  publisher = {Omnipress},
  title = {Fast Newton-type Methods for Total Variation Regularization.},
  url = {http://dblp.uni-trier.de/db/conf/icml/icml2011.html#JimenezS11},
  year = 2011
}

@article{JMLR:v19:13-538,
  author  = {Alvaro Barbero and Suvrit Sra},
  title   = {Modular Proximal Optimization for Multidimensional Total-Variation Regularization},
  journal = {Journal of Machine Learning Research},
  year    = {2018},
  volume  = {19},
  number  = {56},
  pages   = {1-82},
  url     = {http://jmlr.org/papers/v19/13-538.html}
}

@inproceedings{jacot2018neural,
  title={Neural tangent kernel: Convergence and generalization in neural networks},
  author={Jacot, Arthur and Gabriel, Franck and Hongler, Cl{\'e}ment},
  booktitle={Advances in neural information processing systems},
  pages={8571--8580},
  year={2018}
}

@article{heckel2019denoising,
  title={Denoising and Regularization via Exploiting the Structural Bias of Convolutional Generators},
  author={Heckel, Reinhard and Soltanolkotabi, Mahdi},
  journal={arXiv preprint arXiv:1910.14634},
  year={2019}
}

@book{lehoucq1998arpack,
  title={ARPACK users' guide: solution of large-scale eigenvalue problems with implicitly restarted Arnoldi methods},
  author={Lehoucq, Richard B and Sorensen, Danny C and Yang, Chao},
  volume={6},
  year={1998},
  publisher={Siam}
}

@article{honzatko2019accelerating,
  title={Accelerating block-matching and 3D filtering method for image denoising on GPUs},
  author={Honz{\'a}tko, David and Kruli{\v{s}}, Martin},
  journal={Journal of Real-Time Image Processing},
  volume={16},
  number={6},
  pages={2273--2287},
  year={2019},
  publisher={Springer}
}

@article{van2018compressed,
  title={Compressed sensing with deep image prior and learned regularization},
  author={Van Veen, Dave and Jalal, Ajil and Soltanolkotabi, Mahdi and Price, Eric and Vishwanath, Sriram and Dimakis, Alexandros G},
  journal={arXiv preprint arXiv:1806.06438},
  year={2018}
}

@inproceedings{goodfellow2014generative,
  title={Generative adversarial nets},
  author={Goodfellow, Ian and Pouget-Abadie, Jean and Mirza, Mehdi and Xu, Bing and Warde-Farley, David and Ozair, Sherjil and Courville, Aaron and Bengio, Yoshua},
  booktitle={Advances in neural information processing systems},
  pages={2672--2680},
  year={2014}
}

@inproceedings{bora2017compressed,
  title={Compressed sensing using generative models},
  author={Bora, Ashish and Jalal, Ajil and Price, Eric and Dimakis, Alexandros G},
  booktitle={Proceedings of the 34th International Conference on Machine Learning-Volume 70},
  pages={537--546},
  year={2017},
  organization={JMLR. org}
}

@article{daubechies2004iterative,
  title={An iterative thresholding algorithm for linear inverse problems with a sparsity constraint},
  author={Daubechies, Ingrid and Defrise, Michel and De Mol, Christine},
  journal={Communications on Pure and Applied Mathematics: A Journal Issued by the Courant Institute of Mathematical Sciences},
  volume={57},
  number={11},
  pages={1413--1457},
  year={2004},
  publisher={Wiley Online Library}
}

@article{beck2009fast,
  title={A fast iterative shrinkage-thresholding algorithm for linear inverse problems},
  author={Beck, Amir and Teboulle, Marc},
  journal={SIAM journal on imaging sciences},
  volume={2},
  number={1},
  pages={183--202},
  year={2009},
  publisher={SIAM}
}

@article{tibshirani1996regression,
  title={Regression shrinkage and selection via the lasso},
  author={Tibshirani, Robert},
  journal={Journal of the Royal Statistical Society: Series B (Methodological)},
  volume={58},
  number={1},
  pages={267--288},
  year={1996},
  publisher={Wiley Online Library}
}

@article{oymak2019generalization,
  title={Generalization guarantees for neural networks via harnessing the low-rank structure of the jacobian},
  author={Oymak, Samet and Fabian, Zalan and Li, Mingchen and Soltanolkotabi, Mahdi},
  journal={arXiv preprint arXiv:1906.05392},
  year={2019}
}

@inproceedings{liu2019image,
  title={Image restoration using total variation regularized deep image prior},
  author={Liu, Jiaming and Sun, Yu and Xu, Xiaojian and Kamilov, Ulugbek S},
  booktitle={ICASSP 2019-2019 IEEE International Conference on Acoustics, Speech and Signal Processing (ICASSP)},
  pages={7715--7719},
  year={2019},
  organization={IEEE}
}

@inproceedings{mataev2019deepred,
  title={DeepRED: Deep image prior powered by RED},
  author={Mataev, Gary and Milanfar, Peyman and Elad, Michael},
  booktitle={Proceedings of the IEEE International Conference on Computer Vision Workshops},
  pages={0--0},
  year={2019}
}

@article{romano2017little,
  title={The little engine that could: Regularization by denoising (RED)},
  author={Romano, Yaniv and Elad, Michael and Milanfar, Peyman},
  journal={SIAM Journal on Imaging Sciences},
  volume={10},
  number={4},
  pages={1804--1844},
  year={2017},
  publisher={SIAM}
}

@inproceedings{cheng2019bayesian,
  title={A bayesian perspective on the deep image prior},
  author={Cheng, Zezhou and Gadelha, Matheus and Maji, Subhransu and Sheldon, Daniel},
  booktitle={Proceedings of the IEEE Conference on Computer Vision and Pattern Recognition},
  pages={5443--5451},
  year={2019}
}

@article{heckel2018deep,
  title={Deep decoder: Concise image representations from untrained non-convolutional networks},
  author={Heckel, Reinhard and Hand, Paul},
  journal={arXiv preprint arXiv:1810.03982},
  year={2018}
}

@inproceedings{welling2011bayesian,
  title={Bayesian learning via stochastic gradient Langevin dynamics},
  author={Welling, Max and Teh, Yee W},
  booktitle={Proceedings of the 28th international conference on machine learning (ICML-11)},
  pages={681--688},
  year={2011}
}

@article{stein1981estimation,
  title={Estimation of the mean of a multivariate normal distribution},
  author={Stein, Charles M},
  journal={The annals of Statistics},
  pages={1135--1151},
  year={1981},
  publisher={JSTOR}
}

@article{ramani2008monte,
  title={Monte-Carlo SURE: A black-box optimization of regularization parameters for general denoising algorithms},
  author={Ramani, Sathish and Blu, Thierry and Unser, Michael},
  journal={IEEE Transactions on image processing},
  volume={17},
  number={9},
  pages={1540--1554},
  year={2008},
  publisher={IEEE}
}

@article{bojanowski2017optimizing,
  title={Optimizing the latent space of generative networks},
  author={Bojanowski, Piotr and Joulin, Armand and Lopez-Paz, David and Szlam, Arthur},
  journal={arXiv preprint arXiv:1707.05776},
  year={2017}
}

@inproceedings{lai2017deep,
  title={Deep laplacian pyramid networks for fast and accurate super-resolution},
  author={Lai, Wei-Sheng and Huang, Jia-Bin and Ahuja, Narendra and Yang, Ming-Hsuan},
  booktitle={Proceedings of the IEEE conference on computer vision and pattern recognition},
  pages={624--632},
  year={2017}
}

@inproceedings{ledig2017photo,
  title={Photo-realistic single image super-resolution using a generative adversarial network},
  author={Ledig, Christian and Theis, Lucas and Husz{\'a}r, Ferenc and Caballero, Jose and Cunningham, Andrew and Acosta, Alejandro and Aitken, Andrew and Tejani, Alykhan and Totz, Johannes and Wang, Zehan and others},
  booktitle={Proceedings of the IEEE conference on computer vision and pattern recognition},
  pages={4681--4690},
  year={2017}
}

@article{patel2019bayesian,
  title={Bayesian Inference with Generative Adversarial Network Priors},
  author={Patel, Dhruv and Oberai, Assad A},
  journal={arXiv preprint arXiv:1907.09987},
  year={2019}
}

@article{adler2018deep,
  title={Deep bayesian inversion},
  author={Adler, Jonas and {\"O}ktem, Ozan},
  journal={arXiv preprint arXiv:1811.05910},
  year={2018}
}

@inproceedings{ronneberger2015u,
  title={U-net: Convolutional networks for biomedical image segmentation},
  author={Ronneberger, Olaf and Fischer, Philipp and Brox, Thomas},
  booktitle={International Conference on Medical image computing and computer-assisted intervention},
  pages={234--241},
  year={2015},
  organization={Springer}
}

\end{document}